\documentclass[lettersize,journal]{IEEEtran}
\usepackage{amsmath,amssymb,amsfonts}
\usepackage{algorithmic}
\usepackage{array}
\usepackage{cite}
\usepackage[caption=false,font=footnotesize,labelfont=sf,textfont=sf]{subfig}
\usepackage{textcomp}
\usepackage{algorithm}
\usepackage{algorithmic}
\usepackage{stfloats}
\usepackage{url}
\usepackage{verbatim}
\usepackage{graphicx}
\usepackage{xcolor}
\usepackage{bm}
\def\BibTeX{{\rm B\kern-.05em{\sc i\kern-.025em b}\kern-.08em
    T\kern-.1667em\lower.7ex\hbox{E}\kern-.125emX}}
\usepackage{balance}
\begin{document}
\title{Coordinated Multipoint Anti-jamming Beam Pattern Synthesis: From AI Accelerated Algorithm to Hardware Implementation}
\author{Zilong Wang, \textit{Student Member, IEEE}, Cheng Zhang, \textit{Member, IEEE}, Zhilei Zhang, \textit{Member, IEEE},\\ Yaxuan Hu, \textit{Student Member, IEEE}, Wen Wang, \textit{Member, IEEE}, Yongming Huang, \textit{Fellow, IEEE}
\thanks{Zilong Wang, Cheng Zhang, Yaxuan Hu and Yongming Huang are with the National Mobile Communication Research Laboratory, Southeast University, Nanjing 210096, China, and also with the Pervasive Communication Research
Center, Purple Mountain Laboratories, Nanjing 211111, China (e-mail: 
zl{\_}wang@seu.edu.cn, zhangcheng{\_}seu@seu.edu.cn, huyaxuan77@outlook.com, huangym@seu.edu.cn).

Zhilei Zhang and Wen Wang are with the Pervasive Communication Research
Center, Purple Mountain Laboratories, Nanjing 211111, China (e-mail: zhangzhilei@pmlabs.com.cn, wangwen@pmlabs.com.cn). (\textit{Corresponding author}: Cheng Zhang.)}}


\maketitle

\begin{abstract}
This paper presents a deep unfolding-supported coordinated multipoint beam pattern synthesis (DUCoMP-BPS) scheme to overcome the high complexity, poor adaptability, and limited scalability of traditional cell-free anti-jamming beamforming. In the proposed design, access points (APs) independently determine analog beamforming using local angle information, while the central processing unit (CPU) performs cooperative digital beamforming with only a single AP–CPU interaction, significantly reducing fronthaul overhead. To further improve efficiency, a deep unfolding strategy transforms the costly step size search in analog beamforming into a trainable parameter, where an offline-trained complex-valued neural network enables fast and adaptive online inference. Simulation results show that the complexity of DUCoMP-BPS scales linearly with the number of APs, reduces single-AP analog beamforming runtime by about 67\% compared to conventional optimization, and achieves superior nulling performance over purely data-driven approaches. Hardware feasibility is validated on an Advanced RISC Machine–Field Programmable Gate Array (ARM–FPGA) heterogeneous platform, where algorithm–hardware co-verification and hardware–software decoupling enable efficient parallelism and low-latency execution. Finally, anechoic chamber measurements under practical hardware imperfections confirm robust beamforming performance, demonstrating the strong potential of DUCoMP-BPS for real-world deployment.
\end{abstract}


\section{Introduction}
Coordinated multipoint (CoMP) transmission significantly enhances cell-free (CF) communication systems by enabling dynamic coordination among multiple access points (APs) to mitigate the challenges of millimeter-wave (mmWave) propagation, such as severe path loss, blockage effects, and multi-user interference. By jointly optimizing beamforming and resource allocation across distributed APs, CoMP improves spectral efficiency, extends coverage, and enhances interference resilience, thereby boosting overall system performance \cite{10669733}, \cite{10225317}. However, the broadcast nature of wireless channels exposes CF systems to malicious jamming attacks, where adversaries inject high-power interference to disrupt legitimate transmissions. To counter such threats, multiple-input multiple-output (MIMO) beamforming has been widely adopted for anti-jamming transmission, leveraging spatial filtering to nullify jamming signals while maximizing signal-to-interference-plus-noise ratio (SINR) at intended receivers \cite{10422749}.

With large-scale MIMO technology, wireless systems can leverage spatial degrees of freedom to mitigate full-band jamming, which is hard to address via frequency-domain methods. Most anti-jamming research in MIMO systems focuses on beamforming, with varying assumptions regarding the level of knowledge about the jammer\cite{10845845,10699420,9798888}. For example, receive beamforming in the orthogonal subspace of the interference channel was designed in \cite{8781945}, followed by hybrid beamforming at the base station, while a two-layer receive beamforming method was proposed in \cite{9310062}, where the first layer cancels interference via the jamming channel’s orthogonal space and the second layer equalizes the equivalent channel. Such approaches typically require accurate jammer channel state information (CSI), which is often unavailable. To overcome this, the authors of \cite{7952831} considers an uplink single-user MIMO system under jamming during both training and data phases. The interference channel is estimated using unused pilots, enabling receive beamforming and optimized power allocation. Since statistical interference information is easier to obtain than instantaneous CSI, approaches like the Minimum Variance Distortionless Response (MVDR) beamforming have been widely adopted. In \cite{9764295}, MVDR-based fully-digital beamforming was applied to mmWave MIMO, and hybrid beamforming was solved via sparse recovery. To better suit hybrid architectures, the study of \cite{10422749} avoided fully-digital decomposition and introduced alternating optimization for analog and digital beamforming. Recently, reconfigurable intelligent surface (RIS) technology has been explored as a means to enhance spatial anti-jamming capabilities. In \cite{10296481}, RIS was deployed in large-scale MIMO to enable low-cost, energy-efficient beamforming at the user side. The work in \cite{10464890} further extended this by considering hardware constraints such as low-resolution analog-to-digital converters (ADCs). Despite progress, anti-jamming beamforming in CF MIMO systems still faces practical challenges. Specifically, existing methods often involve computationally intensive optimization procedures \cite{10847712}, show poor adaptability to dynamic jamming environments \cite{9686005}, and require extensive CSI exchange among distributed APs, resulting in high signaling overhead and latency.

The vulnerabilities of mmWave systems are further exacerbated by their reliance on line-of-sight (LoS) propagation and sensitivity to hardware impairments (e.g., phase noise, I/Q imbalance) and environmental factors (e.g., user mobility, blockages). Even minor errors in beam alignment, caused by inaccurate user orientation or jamming direction-of-arrival (DoA) prediction, can significantly degrade beamforming gains, thereby undermining communication reliability and anti-jamming effectiveness \cite{10683130}. This issue is critical in high-mobility scenarios where rapid channel variations exceed traditional beam-tracking capabilities, leaving systems vulnerable to persistent jamming attacks. Recent progress in radar-inspired beam pattern synthesis has shown promise in mitigating misalignment issues due to angle estimation errors in mmWave systems \cite{10268603}, \cite{10608392}. These techniques optimize beam characteristics, such as mainlobe, null, and sidelobe, to enhance spatial anti-jamming robustness \cite{8051061}. Various methodologies, including phase-only adjustable phase shifter designs \cite{10268603} and phase-adjustable constraint methods \cite{9611070}, have been proposed to accommodate specific hardware limitations and beamforming requirements. Additionally, beam synthesis techniques in multi-subarray configurations have been explored to further improve spatial selectivity \cite{9142315}, \cite{10223728}. These techniques adjust intra-subarray antenna spacing to produce broader mainlobes and more precise sidelobe control, thereby improving adaptability to complex deployment environments.

However, many of these beam synthesis approaches rely on iterative optimization methods, which are computationally intensive and become impractical for large-scale systems. To alleviate this burden, recent studies have introduced machine learning (ML)-based strategies. For example, \cite{8747423} predicts beamforming weights based on ideal beam pattern parameters, such as mainlobe width, sidelobe levels, and null depth. Similarly, the authors of \cite{9395388} proposed an autoencoder-based solution, where the encoder learns a mapping from desired beam patterns to weights, and the decoder reconstructs the pattern using a global loss function. In \cite{10359475}, a neural network-assisted time-modulated array was proposed, where RF switching times and weights are optimized using an IFFT-based loss function to accelerate beam pattern generation. 

Although data-driven methods substantially reduce computation time, they typically suffer from poor interpretability and limited adaptability, as they often need to be retrained whenever system constraints change \cite{8747423}, \cite{10359475}. To address these limitations, deep unfolding has emerged as a model-driven learning paradigm that integrates domain knowledge into neural networks. This hybrid approach maintains interpretability and adaptability while accelerating optimization processes \cite{9246287}. By learning only a small set of trainable parameters corresponding to the iterations of traditional optimization, deep unfolding significantly reduces computational complexity and improves scalability across diverse scenarios. While several studies \cite{9524496,10286447,Zhang_2020_CVPR} have demonstrated the theoretical advantages of deep unfolding in communication and signal processing tasks, most existing works remain limited to algorithm-level simulations without validating its practical deployment potential. Issues such as hardware compatibility, real-time execution efficiency, and system integration remain largely unexplored.

In light of the aforementioned limitations and to better leverage the potential of mmWave and CoMP-enabled CF systems, we propose a novel deep unfolding-accelerated beam pattern synthesis framework. Unlike prior works that focus solely on algorithmic design, our approach also considers practical hardware implementation. Specifically, we address the challenges of deploying deep unfolding on real-time embedded systems by co-designing an efficient architecture compatible with heterogeneous platforms. The main contributions are as follows.
\begin{itemize}
    \item We propose a deep unfolding-supported coordinated multipoint beam pattern synthesis (DUCoMP-BPS) scheme. In this framework, each AP independently designs its analog beamforming using local angle information, while the CPU jointly optimizes the cooperative digital beamforming. By embedding deep unfolding into the analog design, the step size is treated as a trainable parameter and learned offline via a complex-valued neural network (CvNN). This approach removes the need for iterative step size search during online inference, significantly reducing fronthaul overhead and computational runtime.
    \item The proposed scheme is highly scalable, with complexity increasing only linearly with the number of APs. Simulation results demonstrate that it reduces single-AP analog beamforming runtime by approximately 67\% compared with conventional optimization, while achieving superior null depth relative to purely data-driven methods. For example, in large-scale setups (10 APs × 64 antennas), DUCoMP-BPS achieves mainlobe ripple control within 1 dB, sidelobes below –15 dB, and nulls of –30 dB across 8° angular ranges. Moreover, the scheme exhibits robust anti-jamming communication performance, further confirming its practical advantages.
    \item To validate the hardware feasibility of the proposed algorithm, we design and implement a dedicated architecture on an Advanced RISC Machine–Field Programmable Gate Array (ARM–FPGA) heterogeneous platform. By introducing an algorithm–hardware co-verification chain and adopting a hardware–software decoupled deployment strategy, the system enables reconfigurable adaptation, efficient pipeline parallelism, and hierarchical acceleration of neural network inference, achieving high-performance, low-latency hardware execution.
    \item We validate the proposed algorithm through anechoic chamber measurements. By modeling and compensating for hardware imperfections such as gain/phase inconsistencies and configuration errors, the scheme maintains excellent performance in practice, highlighting its strong potential for real-world deployment.
\end{itemize}

The remainder of the paper is organized as follows. Section II introduces the system model and problem formulation. Section III presents the proposed two-stage anti-jamming beam pattern synthesis with analog/digital beamforming derivation and deep unfolding acceleration. Section IV provides simulation results, Section V details the ARM–FPGA hardware implementation and analysis, Section VI reports practical measurement results, and Section VII concludes the paper.

\textit{Notations}: Use bold lowercase letters to represent column vectors, and bold uppercase letters to represent matrices. The operators $(\cdot)^\mathrm{T}$, $(\cdot)^*$ and $(\cdot)^\mathrm{H}$ correspond to the transpose, conjugate and Hermitian transpose, respectively. The diagonalization and the block diagonalization operations are denoted by $\mathrm{diag}(\cdot)$ and $\mathrm{blkdiag}(\cdot)$, respectively. $\|\mathbf{a}\|$ is the Euclidean norm of the vector $\mathbf{a}$. The symbol $\odot$ represents the Hadamard product. $\mathbb{C}^{M\times N}$ denotes the set of complex-valued $M\times N$ matrices. $\rm{real}(\cdot)$ and $\rm{imag}(\cdot)$ denote the real part and the imaginary part of the vector.

\section{System Model and Problem Formulation}

\subsection{Signal Model}
Consider a millimeter-wave CF massive MIMO system comprising $L$ geographically distributed APs serving $K$ single-antenna user equipments (UEs) in an open area deployment, as shown in Fig. 1. Each AP is equipped with a uniform linear array (ULA)\footnote{Our proposed scheme can also be applied to other antenna arrays in the similar way. In this paper, we mainly investigate the performance of the ULA system.} of $N_r$ antenna elements with a single radio frequency (RF) chain. The array steering vector for the $l$-th AP at direction $\theta$ is expressed as
\begin{equation}
\mathbf{a}_l(\theta) = \left[1, e^{j\frac{2\pi d}{\lambda}\sin\theta}, \cdots, e^{j\frac{2\pi d}{\lambda}(N_r-1)\sin\theta}\right]^{\rm{T}},
\label{eq:steering_vector}
\end{equation}
where $\theta \in [-\pi/2, \pi/2]$ represents the angle of arrival, $d$ denotes the inter-antenna spacing, and $\lambda$ is the carrier wavelength. There are $J$ jammers in the system, and the direction of each jammer can be estimated within a known error range.

\begin{figure}[h!]
\centering
\includegraphics[scale=0.4]{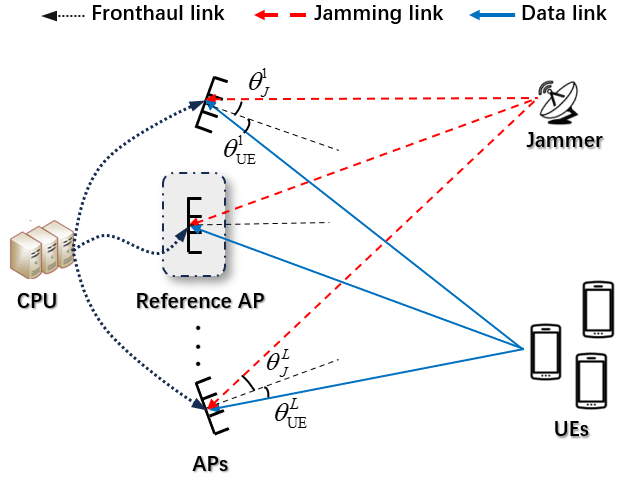}
\caption{Illustration of the considered CoMP CF system.}
\label{fig_1}
\end{figure}

\subsection{Problem Statement}
The LoS dominated propagation characteristic of mmWave systems, while enabling high beamforming gain, introduces significant practical challenges to CF deployments. Practical angular estimation errors lead to severe beam misalignment, which causes received power degradation. Furthermore, the distributed architecture of CF networks makes them vulnerable to sophisticated jamming attacks that exploit beamforming imperfections. Conventional beamforming methods often assume perfect angle information and fixed beam patterns, which fail to ensure reliable communications under realistic conditions. The key challenge is to design a beamforming scheme that can simultaneously tolerate angular errors and suppress jamming, while preserving the spectral efficiency benefits of mmWave CF systems. This requires a joint optimization of beam pattern characteristics under practical constraints and hardware impairments.

\subsection{Problem Formulation}
The joint beamforming design aims to achieve two key objectives, including maintaining stable mainlobe gain for the desired signal despite angle estimation errors and effectively suppressing jamming signals through null steering. This requires minimizing mainlobe ripple while constraining sidelobe levels and creating deep nulls in specified directions. Mathematically, the optimization problem is formulated as
\begin{subequations}\label{eq:optimization}
    \begin{align}
        \label{eq:obj}&\min_{\varepsilon, \mathbf{w}^{\mathrm{BB}}, \mathbf{W}^{\mathrm{RF}}} \varepsilon \\
        \label{eq:mainlobe}\text{s.t.}~&1 - \varepsilon \leq |\mathbf{w}^{\mathrm{BB,H}}\mathbf{W}^{\mathrm{RF,H}}\mathbf{a}(\theta_m)|^2 \leq 1 + \varepsilon, \forall m \in \mathcal{M}, \\
        \label{eq:null}&|\mathbf{w}^{\mathrm{BB,H}}\mathbf{W}^{\mathrm{RF,H}}\mathbf{a}(\theta_n)|^2 \leq \eta_{\mathrm{Z}}, \forall n \in \mathcal{N}, \\
        \label{eq:sidelobe}&|\mathbf{w}^{\mathrm{BB,H}}\mathbf{W}^{\mathrm{RF,H}}\mathbf{a}(\theta_s)|^2 \leq \eta_{\mathrm{SL}}, \forall s \in \mathcal{S}, \\
        \label{eq:analog}&|\mathbf{w}_l^{\mathrm{RF}}(i)| = 1, \forall i \in \mathcal{I},
    \end{align}
\end{subequations}
where $\varepsilon$ represents the mainlobe ripple, $\mathbf{w}^{\mathrm{BB}} \in \mathbb{C}^{L \times 1}$ denotes the digital beamforming vector, and $\mathbf{w}_l^{\mathrm{RF}}\in \mathbb{C}^{N_r \times 1}$ corresponds to the analog beamforming vector for AP $l$. The composite array response vector $\mathbf{a}(\theta_m) = [\mathbf{a}_1^{\mathrm{H}}(\theta_m^1), \cdots, \mathbf{a}_L^{\mathrm{H}}(\theta_m^L)]$ and block-diagonal analog beamforming matrix $\mathbf{W}^{\mathrm{RF}} = \mathrm{blkdiag}(\mathbf{w}_1^{\mathrm{RF}}, \cdots, \mathbf{w}_L^{\mathrm{RF}})$ capture the distributed nature of the CF system. The parameters $\eta_{\mathrm{Z}}$ and $\eta_{\mathrm{SL}}$ control the null depth and sidelobe levels respectively, while $\mathcal{M}$, $\mathcal{N}$, and $\mathcal{S}$ define the angular regions for mainlobe, nulls, and sidelobes. $\mathcal{I}$ is the antenna set of each AP. Constraints \eqref{eq:mainlobe}-\eqref{eq:analog} enforce the mainlobe stability, jamming suppression, and analog beamforming hardware limitations. It is worth mentioning that the angle parameters $\theta_m$, $\theta_n$, $\theta_s$ in \eqref{eq:mainlobe}-\eqref{eq:sidelobe} are those of the reference array in the system. Due to the distinct geometry distribution of the different APs, the local angle parameters of the APs can differ from those of the reference array. Therefore, the angle parameters used in the analog beam design for each AP have superscripts, meaning they are unique and local. The angle parameters of reference array only influence the illustration of beam pattern and have no impact on the performance of the scheme.

The problem in \eqref{eq:optimization} is challenging to solve due to the non-convex coupling between variables, the constant modulus constraint \eqref{eq:analog}, and the high dimensionality of joint digital–analog optimization in distributed systems. Traditional optimization approaches require either excessive fronthaul overhead or high computational capacity of CPU. To address these issues while exploiting the spatial diversity of CF architectures, we propose a novel two-stage beamforming scheme that decouples analog and digital design yet retains coordination gains.

\section{Proposed Two-Stage Scheme Design}

To address the joint beamforming challenge in CF MIMO systems while leveraging their distributed architecture, we propose a novel two-stage optimization framework that decouples the analog and digital beamforming design processes.

\subsection{Two-Stage Optimization Framework}

\subsubsection{Stage 1: Local Analog Beamforming Design}
Each AP independently designs its analog beamformer based on local CSI, by solving the following constrained optimization problem
\begin{subequations}\label{eq:analog_opt}
    \begin{align}
        \label{eq:analog_obj}&\max_{\varepsilon_l, \mathbf{w}_l^{\mathrm{RF}}} \varepsilon_l \\
        \label{eq:analog_main}\text{s.t.}~&\varepsilon_l \leq |\mathbf{w}_l^{\mathrm{RF,H}}\mathbf{a}_l(\theta_m^l)|^2 \leq \alpha \varepsilon_l, \forall m \in \mathcal{M}, \\
        \label{eq:analog_null}&|\mathbf{w}_l^{\mathrm{RF,H}}\mathbf{a}_l(\theta_n^l)|^2 \leq \eta_{\mathrm{Z}}\varepsilon_l, \forall n \in \mathcal{N}, \\
        \label{eq:analog_sidelobe}&|\mathbf{w}_l^{\mathrm{RF,H}}\mathbf{a}_l(\theta_s^l)|^2 \leq \eta_{\mathrm{SL}}\varepsilon_l, \forall s \in \mathcal{S}, \\
        \label{eq:analog_constraint}&|\mathbf{w}_l^{\mathrm{RF}}(i)| = 1, \forall i \in \mathcal{I},
    \end{align}
\end{subequations}
where $\varepsilon_l$ represents the power scaling factor at AP $l$, and $\alpha > 1$ controls the mainlobe ripple tolerance. The constant modulus constraint \eqref{eq:analog_constraint} preserves the analog hardware requirements while the relative gain constraints \eqref{eq:analog_main}-\eqref{eq:analog_sidelobe} maintain the desired beam pattern characteristics.

\subsubsection{Stage 2: Centralized Digital Beamforming Design}
After all APs complete their analog beamformer designs, the CPU collects the composite channel responses and solves the digital beamforming problem
\begin{subequations}\label{eq:digital_opt}
    \begin{align}
        \label{eq:digital_obj}&\min_{\varepsilon, \mathbf{w}^{\mathrm{BB}}} \varepsilon \\
        \label{eq:digital_main}\text{s.t.}~&1 - \varepsilon \leq |\mathbf{w}^{\mathrm{BB,H}}\mathbf{W}^{\mathrm{RF,H}}\mathbf{a}(\theta_m)|^2 \leq 1 + \varepsilon, \forall m \in \mathcal{M}, \\
        \label{eq:digital_null}&|\mathbf{w}^{\mathrm{BB,H}}\mathbf{W}^{\mathrm{RF,H}}\mathbf{a}(\theta_n)|^2 \leq \eta_{\mathrm{Z}}, \forall n \in \mathcal{N}, \\
        \label{eq:digital_sidelobe}&|\mathbf{w}^{\mathrm{BB,H}}\mathbf{W}^{\mathrm{RF,H}}\mathbf{a}(\theta_s)|^2 \leq \eta_{\mathrm{SL}}, \forall s \in \mathcal{S},
    \end{align}
\end{subequations}
where $\mathbf{W}^{\mathrm{RF}} = \mathrm{blkdiag}(\mathbf{w}_1^{\mathrm{RF}}, \cdots, \mathbf{w}_L^{\mathrm{RF}})$ incorporates all APs' analog solutions. This stage fine-tunes the overall beam pattern through digital precoding while respecting the analog hardware constraints.

Both subproblems \eqref{eq:analog_opt} and \eqref{eq:digital_opt} are solved using the Alternating Direction Multiplier Method (ADMM) \cite{MADMM}. This approach efficiently handles the non-convex constraints while maintaining reasonable computational complexity.

\subsection{Analog Beamforming Design at APs}
By using the ADMM approach, sub-problem \eqref{eq:analog_opt} can be transformed to
\begin{subequations}\label{5}
    \begin{align}
\label{5a}&\mathop {\max }\limits_{{\varepsilon _l},{\bf{w}}_l^{{\rm{RF}}}} {\varepsilon _l}\\
\text{s.t.}~\label{5b}&{\varepsilon _l} \le {\left| {{h_m}} \right|^2} \le \alpha {\varepsilon _l},\forall m \in \mathcal{M},\\
\label{5c}&{\left| {{q_n}} \right|^2} \le {\eta _{{\rm{Z}}}}{\varepsilon _l},\forall n \in \mathcal{N},\\
\label{5d}&{\left| {{g_s}} \right|^2} \le {\eta _{{\rm{SL}}}}{\varepsilon _l},\forall s \in \mathcal{S},\\
\label{5e}&\left| {{\bf{w}}_l^{{\rm{RF}}}\left( i \right)} \right| = 1,{\forall i \in {\cal I}},
    \end{align}
\end{subequations}
where ${h_m} = {\bf{w}}_l^{{\rm{RF,H}}}{{\bf{a}}_l}\left( {\theta_m^l} \right)$, ${q_n} = {\bf{w}}_l^{{\rm{RF,H}}}{{\bf{a}}_l}\left( {\theta_n^l} \right)$, and ${g_s} = {\bf{w}}_l^{{\rm{RF,H}}}{{\bf{a}}_l}\left( {\theta_s^l} \right)$. 

Based on the ADMM framework, problem \eqref{5} is transformed to 
\begin{equation}\label{6}
\begin{array}{l}
    \mathop {\min} \limits_{{\varepsilon _l},{\bf{w}}_l^{{\rm{RF}}},{\bf{h}},{\bf{q}},{\bf{g}}} {L\left( {{\bf{w}}_l^{{\rm{RF}}},{\varepsilon _l},{\bf{h}},{\bf{g}},{\bf{q}},{\bm{\delta }},{\bm{\lambda }},{\bm{\xi }}} \right)}\\
    \text{s.t.}~ \eqref{5b},\eqref{5c},\eqref{5d},\eqref{5e},
\end{array}
\end{equation}
where $L$ is the augmented Lagrangian of problem \eqref{5}, i.e.,
\begin{equation}\label{7}
\begin{array}{l}
L\left( {{\bf{w}}_l^{{\rm{RF}}},{\varepsilon _l},{\bf{h}},{\bf{g}},{\bf{q}},{\bm{\delta }},{\bm{\lambda }},{\bm{\xi }}} \right)\\
 =  - {\varepsilon _l} + \frac{\rho }{2}\sum\limits_{m = 1}^M {\left( {{{\left| {{h_m} - {\bf{w}}_l^{{\rm{RF,H}}}{{\bf{a}}_l}\left( {{\theta _m^l}} \right) + {\delta _m}} \right|}^2} - {{\left| {{\delta _m}} \right|}^2}} \right)} \\
 + \frac{\rho }{2}\sum\limits_{s = 1}^S {\left( {{{\left| {{g_s} - {\bf{w}}_l^{{\rm{RF,H}}}{{\bf{a}}_l}\left( {{\theta _s^l}} \right) + {\lambda _s}} \right|}^2} - {{\left| {{\lambda _s}} \right|}^2}} \right)} \\
 + \frac{\rho }{2}\sum\limits_{n = 1}^N {\left( {{{\left| {{q_n} - {\bf{w}}_l^{{\rm{RF,H}}}{{\bf{a}}_l}\left( {{\theta _n^l}} \right) + {\xi _n}} \right|}^2} - {{\left| {{\xi _n}} \right|}^2}} \right)} ,
\end{array}
\end{equation}
where $\rho > 0$ is the penalty parameter. ${\bm{\delta}}=\left \{{\delta_m} \right \}_{m=1}^M$, ${\bm{\lambda}}=\left \{{\lambda_s} \right \}_{s=1}^S$, and ${\bm{\xi}}=\left \{{\xi_n} \right \}_{n=1}^N$ are the Lagrange multipliers. 

Then, problem \eqref{6} can be addressed by dividing it into two independent subproblems and solving them alternately. The first subproblem is as follow
\begin{equation}\label{8}
\begin{array}{l}
\mathop {\min }\limits_{{\bf{w}}_l^{{\rm{RF}}}}~ f({\bf{w}}_l^{{\rm{RF}}})=\left\| {{\bf{u}} - {\bf{w}}_l^{{\rm{RF,H}}}{\bf{A}}} \right\|_2^2~~~\text{s.t.}~\eqref{5e},
\end{array}
\end{equation}
where ${\bf{u}} = {\left[ {{u_{1,1}}, \cdots ,{u_{1,M}},{u_{2,1}}, \cdots ,{u_{2,S}},{u_{3,1}}, \cdots ,{u_{3,N}}} \right]^{\rm{T}}}$, ${u_{1,m}} = {h_m} + {\delta _m}$, 
${u_{2,s}} = {g_s} + {\lambda _s}$, ${u_{3,n}} = {q_n} + {\xi _n}$ and 
${\bf{A}} = [ {{\bf{a}}_l}\left( {{\theta _1^l}} \right), \cdots ,{{\bf{a}}_l}\left( {{\theta _M^l}} \right),{{\bf{a}}_l}\left( {{\theta _1^l}} \right), \cdots ,{{\bf{a}}_l}\left( {{\theta _S^l}} \right), {{\bf{a}}_l}\left( {{\theta _1^l}} \right), \cdots ,$ ${{\bf{a}}_l}\left( {{\theta _N^l}} \right) ]$.

Due to the unitary modulus constraint of analog beamforming, Riemannian gradient descent (RGD) can be a prevalent and useful method to solve the problem. The Euclidean gradient of $f({\bf{w}}_l^{{\rm{RF}}})$ is ${\nabla _{{\bf{w}}_l^{{\rm{RF}}}}}f = {\bf{A}}{{\bf{A}}^{\rm{H}}}{\bf{w}}_l^{{\rm{RF}}} - {\bf{A}}{{\bf{u}}^{\rm{H}}}$. By projecting ${\nabla _{{\bf{w}}_l^{{\rm{RF}}}}}f$ onto the tangent space of the manifold $\mathcal{M}_{cc}$ at the point of ${{\bf{w}}^{\rm RF}_{l}}$, the Riemannian gradient can be expressed as $\nabla _{{\bf{w}}_l^{{\rm{RF}}}}^{\rm{R}}f = {\nabla _{{\bf{w}}_l^{{\rm{RF}}}}}f - {\rm{real}}\left\{ {{\nabla _{{\bf{w}}_l^{{\rm{RF}}}}}f \odot {\bf{w}}_l^{{\rm{RF,*}}}} \right\} \odot {\bf{w}}_l^{{\rm{RF}}}$, where $\mathcal{M}_{cc}= \{{\mathbf{w}} \in {\mathbb{C}}^{N_r}: w^*(i)w(i)=1, \forall i \}$. Then, the iteration equation of ${\bf{w}}_l^{{\rm{RF}}}$ can be expressed as
\begin{equation}\label{9}
    {\bf{w}}_l^{{\rm{RF}}} = \exp \left( {{\rm j}*{\rm angle}\left( {{\bf{w}}_l^{{\rm{RF}}} - \mu \nabla _{{\bf{w}}_l^{{\rm{RF}}}}^{\rm{R}}f} \right)} \right),
\end{equation}
where the operator $\exp(\cdot)$ denotes the exponential function, ${\rm j}$ denotes the imaginary unit, and ${\rm angle}(\cdot)$ denotes the angle of the complex vector. $\mu$ is the step size and can be determined using the Armijo backtracking line search\cite{1966Minimization}.

The second subproblem is expressed as
\begin{equation}\label{10}
\begin{array}{l}
\mathop {\min }\limits_{ {{\varepsilon _l},{\bf{h}},{\bf{g}},{\bf{q}}}}  - {\varepsilon _l} + \frac{\rho }{2}\sum\limits_{m = 1}^M {\left( {{{\left| {{h_m} - {{\hat h}_m}} \right|}^2}} \right)}  + \frac{\rho }{2}\sum\limits_{s = 1}^S {\left( {{{\left| {{g_s} - {{\hat g}_s}} \right|}^2}} \right)}\\
+ \frac{\rho }{2}\sum\limits_{n = 1}^N {\left( {{{\left| {{q_n} - {{\hat q}_n}} \right|}^2}} \right)} \\
\text{s.t.}~\eqref{5b},\eqref{5c},\eqref{5d},
\end{array}  
\end{equation}
where ${\hat h_m} = {\bf{w}}_l^{{\rm{RF,H}}}{{\bf{a}}_l}\left( {{\theta _m^l}} \right) - {\delta _m}$, ${\hat g_s} = {\bf{w}}_l^{{\rm{RF,H}}}{{\bf{a}}_l}\left( {{\theta _s^l}} \right) - {\lambda _s}$ and ${\hat q_n} = {\bf{w}}_l^{{\rm{RF,H}}}{{\bf{a}}_l}\left( {{\theta _n^l}} \right) - {\xi _n}$.

Given ${\varepsilon _l}$, the optimal $h_m$, $g_s$, and $q_n$ can be addressed as
\begin{equation}\label{11}
h_m = \left\{ {\begin{array}{*{20}{c}}
{{{\hat h}_m}}\\
{\sqrt {{\varepsilon _l}} {{\hat h}_m}/\left| {{{\hat h}_m}} \right|}\\
{\sqrt {\alpha {\varepsilon _l}} {{\hat h}_m}/\left| {{{\hat h}_m}} \right|}
\end{array}} \right.\begin{array}{*{20}{c}}
{,\sqrt {{\varepsilon _l}}  \le \left| {{{\hat h}_m}} \right| \le \sqrt {\alpha {\varepsilon _l}} }\\
{,\left| {{{\hat h}_m}} \right| < \sqrt {{\varepsilon _l}} }\\
{,\left| {{{\hat h}_m}} \right| > \sqrt {\alpha {\varepsilon _l}} },
\end{array}
\end{equation}
\begin{equation}\label{12}
g_s = \left\{ {\begin{array}{*{20}{c}}
{{{\hat g}_s}}&{,\left| {{{\hat g}_s}} \right| \le \sqrt {{\eta _{SL}}{\varepsilon _l}} }\\
{\sqrt {{\eta _{SL}}{\varepsilon _l}} {{\hat g}_s}/\left| {{{\hat g}_s}} \right|}&{,otherwise},
\end{array}} \right.
\end{equation}
\begin{equation}\label{13}
q_n = \left\{ {\begin{array}{*{20}{c}}
{{{\hat q}_n}}&{,\left| {{{\hat q}_n}} \right| \le \sqrt {{\eta _{Z}}{\varepsilon _l}} }\\
{\sqrt {{\eta _{Z}}{\varepsilon _l}} {{\hat q}_n}/\left| {{{\hat q}_n}} \right|}&{,otherwise}.
\end{array}} \right.
\end{equation}

Applying \eqref{11}-\eqref{13} in \eqref{10}, we can obtain the subproblem with respect to $\sqrt{\varepsilon _l}$ as 
\begin{equation}\label{14}
\begin{array}{l}
\mathop {\min }\limits_{{\varepsilon _l}} g\left( {\sqrt{\varepsilon _l}} \right) =  - {\varepsilon _l} + \sum\limits_{m = 1}^M {{{\left( {\sqrt {\alpha {\varepsilon _l}}  - \left| {{{\hat h}_m}} \right|} \right)}^2}}  \\+ \sum\limits_{m = 1}^M {{{\left( {\sqrt {{\varepsilon _l}}  - \left| {{{\hat h}_m}} \right|} \right)}^2}}  + \sum\limits_{s = 1}^S {{{\left( {\sqrt {{\eta _{SL}}{\varepsilon _l}}  - \left| {{{\hat g}_s}} \right|} \right)}^2}} 
\\ + \sum\limits_{n = 1}^N {{{\left( {\sqrt {{\eta _{Z}}{\varepsilon _l}}  - \left| {{{\hat q}_n}} \right|} \right)}^2}} .
\end{array}
\end{equation}
To solve this problem, we sort $\left\{ {\left| {{{\hat h}_m}} \right|} \right\}_{m=1}^{M}$, $\left\{ {\left| {{{\hat h}_m}} \right|}/\sqrt{\alpha} \right\}_{m=1}^{M}$, $\left\{ {\left| {{{\hat g}_s}} \right|} \right\}$, and $\left\{ {\left| {{{\hat q}_n}} \right|} \right\}$ in an ascending order with duplicates removed as $\mathbf{r}$, $\mathbf{v}$, ${\mathbf{r'}}$, and $\mathbf{v'}$, respectively. And we sort all the elements in $\mathbf{r}$, $\mathbf{v}$, ${\mathbf{r'}}$, and $\mathbf{v'}$ in an ascending order as $\mathbf{e}$. Then, $g(\sqrt{\varepsilon _l})$ can be expressed as a piecewise function
\begin{equation}\label{15}
g\left( {{\varepsilon _l}} \right) = \left\{ {{g_j}\left( {\sqrt {{\varepsilon _l}} } \right)|{e_{j - 1}} \le \sqrt {{\varepsilon _l}}  \le {e_j},j = 2, \cdots ,J} \right\},
\end{equation}
where ${g_j}\left( {\sqrt {{\varepsilon _l}} } \right) = {a_j}{\varepsilon _l} + {b_j}\sqrt {{\varepsilon _l}}  + {c_j}$ and
\begin{equation}\label{16}
\begin{array}{l}
{a_j} = \left( {\sum\limits_{i = m''}^{M - 1} \alpha   + \sum\limits_{i = 1}^{m'} 1  + \sum\limits_{i = 1}^{s'} {{\eta _{SL}}}  + \sum\limits_{i = 1}^{n'} {{\eta _{Z,l}}} } \right) - 1,\\
{b_j} =  - \left( {\sum\limits_{i = m''}^{M - 1} {\sqrt \alpha  {v_i}}  + \sum\limits_{i = 1}^{m'} {r{}_i + \sum\limits_{i = 1}^{s'} {{{r'}_i}} }  + \sum\limits_{i = 1}^{n'} {{{v'}_i}} } \right),\\
{c_j} = \left( {\sum\limits_{i = m''}^{M - 1} {v_i^2}  + \sum\limits_{i = 1}^{m'} {r_i^2 + \sum\limits_{i = 1}^{s'} {\left( {r'} \right)_i^2} }  + \sum\limits_{i = 1}^{n'} {\left( {{{v'}_i}} \right)_i^2} } \right),
\end{array}
\end{equation}
where $m'$, $m''$, $s'$, $n'$ satisfy $\left[ {{e_{j - 1}},{e_j}} \right] \subseteq \left[ {{r_{m'}},{r_{m' + 1}}} \right]$, $\left[ {{e_{j - 1}},{e_j}} \right] \subseteq \left[ {{v_{m'' - 1}},{v_{m''}}} \right]$, $\left[ {{e_{j - 1}},{e_j}} \right] \subseteq \left[ {{{r'}_{s' - 1}},{{r'}_{s'}}} \right]$ and $\left[ {{e_{j - 1}},{e_j}} \right] \subseteq \left[ {{{v'}_{n' - 1}},{{v'}_{n'}}} \right]$, respectively. Obviously, ${g_j}\left( {\sqrt {{\varepsilon _l}} } \right)$ is a quadratic function, the potential minimal value can be obtained when ${\left( {\sqrt {{\varepsilon _l}} } \right)_j} = \arg \min\limits_{{\varepsilon _l}} \left\{ {{g_j}\left( {{e_{j - 1}}} \right),{g_j}\left( {{e_j}} \right),{g_j}\left( { - {{{b_j}} \mathord{\left/ {\vphantom {{{b_j}} {2{a_j}}}} \right.} {2{a_j}}}} \right)} \right\}$. By collecting all the potential minimal values of all the pieces, we can easily get the optimal value of $\varepsilon _l$
\begin{equation}\label{17}
{{ {{\varepsilon _l}} }} = \left \{ \arg \min\limits_{{{\left( {\sqrt {{\varepsilon _l}} } \right)}_j}} \left\{ {{g_1}\left( {{{\left( {\sqrt {{\varepsilon _l}} } \right)}_1}} \right), \cdots ,{g_J}\left( {{{\left( {\sqrt {{\varepsilon _l}} } \right)}_J}} \right)} \right\} \right \}^2.
\end{equation}

After determining the optimal variables, the Lagrangian multipliers are updated using the following rules:
\begin{equation}\label{18}
\begin{array}{l}
{\delta _m} = {\delta _m} + \rho \left( {{h_m} - {\bf{w}}_l^{{\rm{RF,H}}}{{\bf{a}}_l}\left( {{\theta _m^l}} \right)} \right),\\
{\lambda _s} = {\lambda _s} + \rho \left( {{g_s} - {\bf{w}}_l^{{\rm{RF,H}}}{{\bf{a}}_l}\left( {{\theta _s^l}} \right)} \right),\\
{\xi _n} = {\xi _n} + \rho \left( {{q_n} - {\bf{w}}_l^{{\rm{RF,H}}}{{\bf{a}}_l}\left( {{\theta _n^l}} \right)} \right).
\end{array}
\end{equation}

The analog beamforming scheme of each AP is summarized as \textbf{Algorithm \ref{alg1}}. 

\begin{algorithm}[h!]
{{
\caption{Analog Beamforming Design of each AP.}
\label{alg1}
\begin{algorithmic}[1]
\STATE{\textbf{Input}: ${\eta _{\rm{Z}}}$, ${\eta _{\rm{SL}}}$, ${{\theta _m^l}}$, ${{\theta _s^l}}$, ${{\theta _n^l}}$}
\STATE{\textbf{Initialize}: ${{\bf{w}}_l^{{\rm{RF}}},{\varepsilon _l},{\bf{h}},{\bf{g}},{\bf{q}},{\bm{\delta }},{\bm{\lambda }},{\bm{\xi }}}$ }
\STATE{${\mathbf{while}}$ $(iter \le itermax~\mathbf{or}~\varepsilon^{t+1}_{l}-\varepsilon^{t}_{l}\le \kappa)$}
\STATE \hspace{0.5cm}{Calculate ${{{\bf{w}}}^{\rm RF}_{l}}$ by \eqref{9} iteratively}
\STATE \hspace{0.5cm}{Calculate ${{\bf{h}}},{{\bf{g}}},{{\bf{q}}}$ by \eqref{11}-\eqref{13} and calculate $\varepsilon_{l}$ by \eqref{17}}
\STATE \hspace{0.5cm}{Calculate ${{\bm{\delta }}},{{\bm{\lambda }}},{{\bm{\xi }}}$ by \eqref{18}}
\STATE {\textbf{Return}: ${{{\bf{w}}}^{\rm RF}_{l}}$}
\end{algorithmic}
}}
\end{algorithm}

\subsection{Digital Beamforming Design at CPU}\label{AA}
Upon obtaining the optimized analog beamformers from all APs, the CPU proceeds to compute the optimal digital beamforming solution by solving sub-problem \eqref{eq:digital_opt}. Following a similar methodology to the analog beamforming solution, we first reformulate the digital beamforming problem as
\begin{subequations}\label{19}
\begin{align}
\label{19a}&{\mathop {\min }\limits_{\varepsilon ,{{\bf{w}}^{{\rm{BB}}}}} \varepsilon }\\
\text{s.t.}~\label{19b}&{1 - \varepsilon  \le {{\left| {{h^{\rm cpu}_m}} \right|}^2} \le 1 + \varepsilon ,\forall m \in {\cal M}},\\
\label{19c}&{{{\left| {{q^{\rm cpu}_n}} \right|}^2} \le {\eta _{\rm{Z}}},\forall n \in {\cal N}},\\
\label{19d}&{{{\left| {{g^{\rm cpu}_s}} \right|}^2} \le {\eta _{{\rm{SL}}}},\forall s \in {\cal S}},
\end{align}
\end{subequations}
where $h_m^{{\rm{cpu}}} = {{\bf{w}}^{{\rm{BB}},{\rm{H}}}}{{\bf{a}}_{{\rm{cpu}}}}\left( {{\theta _m}} \right)$, $g_s^{{\rm{cpu}}} = {{\bf{w}}^{{\rm{BB}},{\rm{H}}}}{{\bf{a}}_{{\rm{cpu}}}}\left( {{\theta _s}} \right)$, $q_n^{{\rm{cpu}}} = {{\bf{w}}^{{\rm{BB}},{\rm{H}}}}{{\bf{a}}_{{\rm{cpu}}}}\left( {{\theta _n}} \right)$ and ${{\bf{a}}_{{\rm{cpu}}}}\left( {{\theta _i}} \right) = {{\bf{W}}^{{\rm{RF}},{\rm{H}}}}{\bf{a}}\left( {{\theta _i}} \right)$. Similarly, problem \eqref{19} is transformed to 
\begin{equation}\label{20}
    \begin{array}{l}
    \mathop {\min} \limits_{\scriptstyle {{\bf{w}}^{{\rm{BB}}}},\varepsilon, {{\bf{h}}^{{\rm{cpu}}}},\hfill\atop \scriptstyle{{\bf{g}}^{{\rm{cpu}}}},{{\bf{q}}^{{\rm{cpu}}}}} {L\left( {{{\bf{w}}^{{\rm{BB}}}},\varepsilon ,{{\bf{h}}^{{\rm{cpu}}}},{{\bf{g}}^{{\rm{cpu}}}},{{\bf{q}}^{{\rm{cpu}}}},{{\bm{\delta }}^{{\rm{cpu}}}},{{\bm{\lambda }}^{{\rm{cpu}}}},{{\bm{\xi }}^{{\rm{cpu}}}}} \right)}\\
    \text{s.t.}~ \eqref{19b},\eqref{19c},\eqref{19d}
\end{array}
\end{equation}
where $L$ is the augmented Lagrangian, i.e.,
\begin{equation}\label{21}
\begin{array}{l}
L\left( {{{\bf{w}}^{{\rm{BB}}}},\varepsilon ,{{\bf{h}}^{{\rm{cpu}}}},{{\bf{g}}^{{\rm{cpu}}}},{{\bf{q}}^{{\rm{cpu}}}},{{\bm{\delta }}^{{\rm{cpu}}}},{{\bm{\lambda }}^{{\rm{cpu}}}},{{\bm{\xi }}^{{\rm{cpu}}}}} \right)\\
 = \varepsilon  + \frac{\rho }{2}\sum\limits_{m = 1}^M {\left( {{{\left| {h_m^{{\rm{cpu}}} - {\bf{w}}_{}^{{\rm{BB,H}}}{{\bf{a}}_{{\rm{cpu}}}}\left( {{\theta _m}} \right) + \delta _m^{{\rm{cpu}}}} \right|}^2} - {{\left| {\delta _m^{{\rm{cpu}}}} \right|}^2}} \right)} \\
 + \frac{\rho }{2}\sum\limits_{s = 1}^S {\left( {{{\left| {g_s^{{\rm{cpu}}} - {\bf{w}}_{}^{{\rm{BB,H}}}{{\bf{a}}_{{\rm{cpu}}}}\left( {{\theta _s}} \right) + \lambda _s^{{\rm{cpu}}}} \right|}^2} - {{\left| {\lambda _s^{{\rm{cpu}}}} \right|}^2}} \right)} \\ + \frac{\rho }{2}\sum\limits_{n = 1}^N {\left( {{{\left| {q_n^{{\rm{cpu}}} - {\bf{w}}_{}^{{\rm{BB,H}}}{{\bf{a}}_{{\rm{cpu}}}}\left( {{\theta _n}} \right) + \xi _n^{{\rm{cpu}}}} \right|}^2} - {{\left| {\xi _n^{{\rm{cpu}}}} \right|}^2}} \right)} ,
\end{array}
\end{equation}
where $\rho > 0$ is the penalty parameter. ${\bm{\delta}^{\rm cpu}}=\left \{{\delta^{\rm cpu}_m} \right \}_{m=1}^M$, ${\bm{\lambda}^{\rm cpu}}=\left \{{\lambda^{\rm cpu}_s} \right \}_{s=1}^S$ and ${\bm{\xi}^{\rm cpu}}=\left \{{\xi^{\rm cpu}_n} \right \}_{n=1}^N$ are the Lagrange multipliers.

Similar to \eqref{6}, problem \eqref{20} can be addressed by dividing it into two independent subproblems and solving them alternately. The first subproblem is as follow
\begin{equation}\label{22}
\mathop {\min }\limits_{{\bf{w}}^{{\rm{BB,H}}}} \left\| {{{\bf{u}}_{{\rm{cpu}}}} - {\bf{w}}^{{\rm{BB,H}}}{{\bf{A}}_{{\rm{cpu}}}}} \right\|_2^2,
\end{equation}
where ${{\bf{A}}_{{\rm{cpu}}}} = [ {{\bf{a}}_{{\rm{cpu}}}}\left( {{\theta _1}} \right), \cdots ,{{\bf{a}}_{{\rm{cpu}}}}\left( {{\theta _M}} \right),{{\bf{a}}_{{\rm{cpu}}}}\left( {{\theta _1}} \right), \cdots , $ ${{\bf{a}}_{{\rm{cpu}}}}\left( {{\theta _S}} \right),{{\bf{a}}_{{\rm{cpu}}}}\left( {{\theta _1}} \right), \cdots ,{{\bf{a}}_{{\rm{cpu}}}}\left( {{\theta _N}} \right) ]$, ${{\bf{u}}_{{\rm{cpu}}}} = [ u_{1,1}^{{\rm{cpu}}}, \cdots ,$ $u_{1,M}^{{\rm{cpu}}},u_{2,1}^{{\rm{cpu}}}, \cdots ,u_{2,S}^{{\rm{cpu}}},u_{3,1}^{{\rm{cpu}}}, \cdots ,u_{3,N}^{{\rm{cpu}}}]^{\rm{T}}$, $u_{1,m}^{{\rm{cpu}}} = h_m^{{\rm{cpu}}} + \delta _m^{{\rm{cpu}}}$, $u_{2,s}^{{\rm{cpu}}} = g_s^{{\rm{cpu}}} + \lambda _s^{{\rm{cpu}}}$, $u_{3,n}^{{\rm{cpu}}} = q_n^{{\rm{cpu}}} + \xi _n^{{\rm{cpu}}}$. Least-squares method is convenient to obtain the optimal solution of this problem, and the closed form of the solution is
\begin{equation}\label{23}
{\bf{w}}^{{\rm{BB}}} = {\left( {{{\bf{A}}_{{\rm{cpu}}}}{\bf{A}}_{{\rm{cpu}}}^{\rm{H}}} \right)^{ - 1}}{{\bf{A}}_{{\rm{cpu}}}}{\bf{u}}_{{\rm{cpu}}}^{\rm{H}}.
\end{equation}

Then, the second subproblem is 
\begin{equation}\label{24}
\begin{array}{l}
\mathop {\min }\limits_{\varepsilon ,{{\bf{h}}^{{\rm{cpu}}}},{{\bf{g}}^{{\rm{cpu}}}},{{\bf{q}}^{{\rm{cpu}}}}} \varepsilon  + \frac{\rho }{2}\sum\limits_{m = 1}^M {\left( {{{\left| {h_m^{{\rm{cpu}}} - \hat h_m^{{\rm{cpu}}}} \right|}^2}} \right)}  \\+ \frac{\rho }{2}\sum\limits_{s = 1}^S {\left( {{{\left| {g_s^{{\rm{cpu}}} - \hat g_s^{{\rm{cpu}}}} \right|}^2}} \right)}  + \frac{\rho }{2}\sum\limits_{n = 1}^N {\left( {{{\left| {q_n^{{\rm{cpu}}} - \hat q_n^{{\rm{cpu}}}} \right|}^2}} \right)} \\
\text{s.t.}~ \eqref{19b}, \eqref{19c}, \eqref{19d},
\end{array}
\end{equation}
where $\hat h_m^{{\rm{cpu}}} = {\bf{w}}_{}^{{\rm{BB,H}}}{{\bf{a}}_{{\rm{cpu}}}}\left( {{\theta _m}} \right) - \delta _m^{{\rm{cpu}}}$, $\hat g_s^{{\rm{cpu}}} = {\bf{w}}_{}^{{\rm{BB,H}}}{{\bf{a}}_{{\rm{cpu}}}}\left( {{\theta _s}} \right) - \lambda _s^{{\rm{cpu}}}$ and $\hat q_n^{{\rm{cpu}}} = {\bf{w}}_{}^{{\rm{BB,H}}}{{\bf{a}}_{{\rm{cpu}}}}\left( {{\theta _n}} \right) - \xi _n^{{\rm{cpu}}}$.

Given $\varepsilon$, the optimal $h_m^{\rm cpu}$, $g_s^{\rm cpu}$ and $q_n^{\rm cpu}$ can be addressed as
\begin{equation}\label{25}
h_m^{{\rm{cpu}}} = \left\{ {\begin{array}{*{20}{c}}
{\hat h_m^{{\rm{cpu}}}}\\
{\sqrt {1 - \varepsilon } \hat h_m^{{\rm{cpu}}}/\left| {\hat h_m^{{\rm{cpu}}}} \right|}\\
{\sqrt {1 + \varepsilon } \hat h_m^{{\rm{cpu}}}/\left| {\hat h_m^{{\rm{cpu}}}} \right|}
\end{array}} \right.\begin{array}{*{20}{c}}
{,\sqrt {1 - \varepsilon }  \le \left| {\hat h_m^{{\rm{cpu}}}} \right| \le \sqrt {1 + \varepsilon } }\\
{,\left| {\hat h_m^{{\rm{cpu}}}} \right| < \sqrt {1 - \varepsilon } }\\
{,\left| {\hat h_m^{{\rm{cpu}}}} \right| > \sqrt {1 + \varepsilon } },
\end{array}
\end{equation}
\begin{equation}\label{26}
g_s^{{\rm{cpu}}} = \left\{ {\begin{array}{*{20}{c}}
{\hat g_s^{{\rm{cpu}}}}&{,\left| {\hat g_s^{{\rm{cpu}}}} \right| \le \sqrt {{\eta _{SL}}} }\\
{\sqrt {{\eta _{SL}}} \hat g_s^{{\rm{cpu}}}/\left| {\hat g_s^{{\rm{cpu}}}} \right|}&{,otherwise},
\end{array}} \right.
\end{equation}
\begin{equation}\label{27}
q_n^{{\rm{cpu}}} = \left\{ {\begin{array}{*{20}{c}}
{\hat q_n^{{\rm{cpu}}}}&{,\left| {\hat q_n^{{\rm{cpu}}}} \right| \le \sqrt {{\eta _Z}} }\\
{\sqrt {{\eta _Z}} \hat q_n^{{\rm{cpu}}}/\left| {\hat q_n^{{\rm{cpu}}}} \right|}&{,otherwise}.
\end{array}} \right.
\end{equation}

Then we sort $\left\{ {\sqrt {\left| {1 - {{\left| {h_m^{{\rm{cpu}}}} \right|}^2}} \right|} } \right\}_{m = 1}^M$ in the ascending order as $\mathbf{e}$. Similarly, we can obtain the problem with respect to $\varepsilon$
\begin{equation}\label{28}
\mathop {\min }\limits_\varepsilon  h\left( \varepsilon  \right) = \varepsilon  + \sum\limits_{m = 1}^M {{{\left( {\sqrt \varepsilon   - {e_m}} \right)}^2}} .
\end{equation}
It is convenient to get the optimal solution of $\varepsilon$
\begin{equation}\label{29}
\varepsilon  = \left\{ {\begin{array}{*{20}{c}}
{{{\left| {{e_m}} \right|}^2},}&{\left| {{e_m}} \right| \ge \frac{{{{\left\| {\bf{e}} \right\|}_1}}}{{M + 1}}}\\
{{{\left| {{e_{m + 1}}} \right|}^2},}&{\left| {{e_{m + 1}}} \right| \le \frac{{{{\left\| {\bf{e}} \right\|}_1}}}{{M + 1}}}\\
{{{\left( {\frac{{{{\left\| {\bf{e}} \right\|}_1}}}{{M + 1}}} \right)}^2},}&{otherwise}.
\end{array}} \right.
\end{equation}

After determining the optimal variables of CPU, the Lagrangian multipliers are updated using the following rules:
\begin{equation}\label{30}
\begin{array}{l}
\delta _m^{{\rm{cpu}}} = \delta _m^{{\rm{cpu}}} + \rho \left( {h_m^{{\rm{cpu}}} - {\bf{w}}^{{\rm{BB,H}}}{{\bf{a}}_{{\rm{cpu}}}}\left( {{\theta _m}} \right)} \right),\\
\lambda _s^{{\rm{cpu}}} = \lambda _s^{{\rm{cpu}}} + \rho \left( {g_s^{{\rm{cpu}}} - {\bf{w}}^{{\rm{BB,H}}}{{\bf{a}}_{{\rm{cpu}}}}\left( {{\theta _s}} \right)} \right),\\
\xi _n^{{\rm{cpu}}} = \xi _n^{{\rm{cpu}}} + \rho \left( {q_n^{{\rm{cpu}}} - {\bf{w}}^{{\rm{BB,H}}}{{\bf{a}}_{{\rm{cpu}}}}\left( {{\theta _n}} \right)} \right).
\end{array}
\end{equation}

\subsection{Deep Unfolding Design}
The proposed two-stage scheme provides robust anti-jamming beamforming for CF MIMO systems. While most subproblems admit closed-form solutions, subproblem \eqref{8} requires iterative RGD due to the unitary modulus constraint. Traditional RGD relies on Armijo backtracking line search to determine the step size, which slows convergence and makes it unsuitable for real-time operation when user or jammer directions vary rapidly. 

\begin{figure}[h!]
\centering
\includegraphics[scale=0.30]{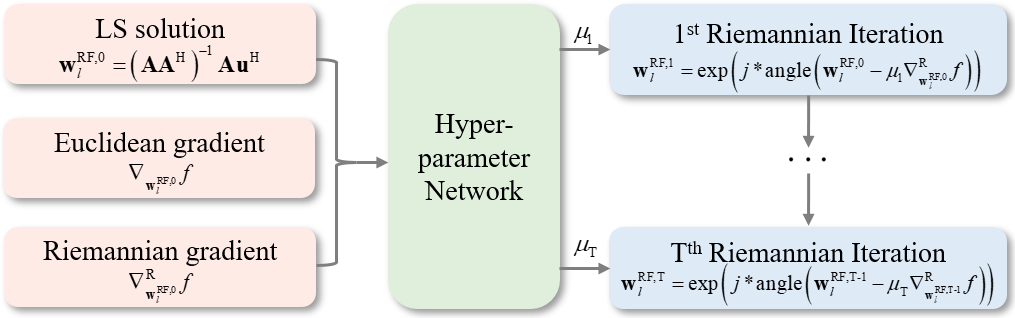}
\caption{Architecture of deep unfolding RGD with hyperparameter network}
\label{fig_2}
\end{figure}

To address this limitation, we adopt a deep unfolding approach, a model-based machine learning technique that embeds domain knowledge into a neural network to accelerate iterative optimization. Specifically, a CvNN is trained to predict the optimal step size for each RGD iteration in problem \eqref{8}. As shown in Fig. \ref{fig_2}, the architecture consists of:

\begin{itemize}
\item A hyperparameter network with five linear CvNN layers utilizing complex rectified linear unit (CReLU) activations (except the output layer).
\item A final layer that combines real and imaginary parts with absolute value activation to ensure non-negative step sizes.
\item $T$ unfolding layers, each representing one Riemannian iteration.
\end{itemize}

The network takes three key inputs:
\begin{enumerate}
\item The least squares solution ${\left( {{\bf{AA}}^{\rm{H}}} \right)^{ - 1}}{\bf{Au}}^{\rm{H}}$ of problem \eqref{8}.
\item The corresponding Euclidean gradient $\nabla _{{\bf{w}}_l^{{\rm{RF,0}}}}f$.
\item The corresponding Riemannian gradient $\nabla _{{\bf{w}}_l^{{\rm{RF,0}}}}^{\rm{R}}f$.
\end{enumerate}
and outputs optimized step sizes ${\mu_{t}}$ for each RGD iteration. We initialize the analog beamforming as
\begin{equation}
{\bf{w}}_l^{{\rm{RF,0}}} = {\left( {{\bf{AA}}^{\rm{H}}} \right)^{ - 1}}{\bf{Au}}^{\rm{H}}.
\end{equation}

The training process uses the sum of squared errors from \eqref{8} as the loss function:
\begin{equation}\label{31}
\mathcal{L}=\frac{1}{T}\sum\limits_{t = 1}^T {\left\| {{{\bf{u}}_t} - {\bf{w}}_{l,t}^{{\rm{RF,H}}}{{\bf{A}}_t}} \right\|_2^2},
\end{equation}
where $T$ is the maximum iteration count. The training set is generated with randomized inputs to ensure generalization. Once trained, the hyperparameter network can be applied to all APs without retraining, as their analog beamforming problems share the same structure. This approach removes the need for Armijo line search in Step of Algorithm \ref{alg1}, thereby significantly reducing computation time. The proposed deep unfolding accelerated CoMP anti-jamming beam pattern synthesis scheme is summarized in \textbf{Algorithm \ref{alg1}}.

\begin{algorithm}[h!]
{{
\caption{Deep Unfolding Supported CoMP Beam Pattern Synthesis Design (DUCoMP-BPS).}
\label{alg2}
\begin{algorithmic}[1]
\STATE{\textbf{Input}: ${\eta _{\rm{Z}}}$, ${\eta _{\rm{SL}}}$, ${{\theta _m}}$, ${{\theta _s}}$, ${{\theta _n}}$}
\STATE{\textbf{Initialize}: ${{{\bf{W}}}^{\rm RF}}$, ${{{\bf{w}}}^{\rm BB}}$, ${\varepsilon}$, ${{\bf{h}}^{{\rm{cpu}}}}$, ${{\bf{g}}^{{\rm{cpu}}}}$, ${{\bf{q}}^{{\rm{cpu}}}}$, ${{\bm{\delta }}^{{\rm{cpu}}}}$, ${{\bm{\lambda }}^{{\rm{cpu}}}}$, ${{\bm{\xi }}^{{\rm{cpu}}}}$}
\STATE{${\mathbf{while}}$ $(iter \le itermax~\mathbf{or}~\varepsilon^{t+1}-\varepsilon^{t}\le \kappa)$}
\STATE \hspace{0.5cm}{Calculate ${{{\bf{w}}}^{\rm RF}_{l}}$ of AP $l$ by deep unfolding modified \textbf{Algorithm 1} in parallel }
\STATE \hspace{0.5cm}{Calculate ${{{\bf{w}}}^{\rm BB}}$ by \eqref{23}}
\STATE \hspace{0.5cm}{Calculate ${{\bf{h}}^{{\rm{cpu}}}},{{\bf{g}}^{{\rm{cpu}}}},{{\bf{q}}^{{\rm{cpu}}}}$ by \eqref{25}, \eqref{26}, \eqref{27} and then calculate $\varepsilon$ by \eqref{29}}
\STATE \hspace{0.5cm}{Calculate ${{\bm{\delta }}^{{\rm{cpu}}}},{{\bm{\lambda }}^{{\rm{cpu}}}},{{\bm{\xi }}^{{\rm{cpu}}}}$ by \eqref{30}}
\STATE {\textbf{Return}: ${{{\bf{W}}}^{\rm RF}}$, ${{{\bf{w}}}^{\rm BB}}$}
\end{algorithmic}
}}
\end{algorithm}

\subsection{Complexity and Convergence Analysis}\label{complex_ref}

For \textbf{Algorithm \ref{alg1}}, the complexity is $\mathcal{O}(T(M+N+S)N_r^2)$, where $T$ represents the number of RGD iterations. For \textbf{Algorithm \ref{alg2}}, the total complexity is $\mathcal{O}(L^3 + L^2(M+N+S) + L(M+N+S)^2 + T(M+N+S)N_r^2)$. In practical CF MIMO deployments, , the number of APs L is typically much smaller than the total number of angular sampling points $M+N+S$ (e.g., typical values being $N_r = 64$, $M+N+S = 180$, and $L = 10$). Under this setting, the dominant terms reduce the complexity to $\mathcal{O}(L(M+N+S)^2 + T(M+N+S)N_r^2)$. This expression demonstrates that as the number of APs increases, the total computational complexity grows linearly, which confirms the scalability of the proposed scheme for large-scale CF systems.

For convergence analysis, we focus on the analog beamforming problem, as the digital beamforming case follows similar principles. Regarding convergence properties, the RGD method for analog beamforming exhibits linear convergence to a local optimum, guaranteed by the smoothness of the objective function on the complex circle manifold $\mathcal{M}_{cc}$. The ADMM-based scheme in \textbf{Algorithm \ref{alg1}} converges to a stationary point when the following conditions are met: $\lim_{t \to \infty} \|\bm{\delta}^{t+1} - \bm{\delta}^{t}\| = 0$, $\lim_{t \to \infty} \|\bm{\lambda}^{t+1} - \bm{\lambda}^{t}\| = 0$, and $\lim_{t \to \infty} \|\bm{\xi}^{t+1} - \bm{\xi}^{t}\| = 0$. These conditions ensure that both \textbf{Algorithm \ref{alg1}} and \textbf{Algorithm \ref{alg2}} converge to stable solutions. Moreover, the convergence characteristics remain consistent across different problem sizes, which further validates the practicality and robustness of the proposed scheme.

\section{Simulation-Based Performance Evaluation}
This section presents a comprehensive simulation-based evaluation of the proposed DUCoMP-BPS scheme. The objective is threefold: (i) to verify its anti-jamming and beam pattern performance under various scenarios, (ii) to quantify computational complexity and scalability, and (iii) to generate design insights that will guide the hardware acceleration architecture in Section V. Unless otherwise stated, all simulations assume that the UEs and jammer satisfy the far-field condition for each AP. We set $L=10$ and $N_r=64$ for the system and $itermax=50$, $\rho=1e-5$, $\alpha=1.05$, $\eta_{SL}=-15dB$ and $\eta_{Z}=-30dB$ for both algorithms. The hyperparameter network is trained with a batch size of 100 and an output size $T=15$.

\subsection{Analog Beam Pattern Performance}
First, we evaluate the proposed deep unfolding aided analog beam pattern performance of single AP. We compare the proposed deep unfolding scheme with different benchmarks:
\begin{itemize}
    \item {\bf{Low Sidelobe}}: method in literature \cite{9611070}. 
    \item {\bf{Riemannian}}: the proposed scheme without deep unfolding acceleration.
    \item {\bf{Black Box}}: a black box composed of a 3-layer deep neural network (DNN), in which the input is the ideal beam pattern and the output is the angle of weights of the analog beamforming.
\end{itemize}

\begin{figure}[h]
\centering
\includegraphics[scale=0.5]{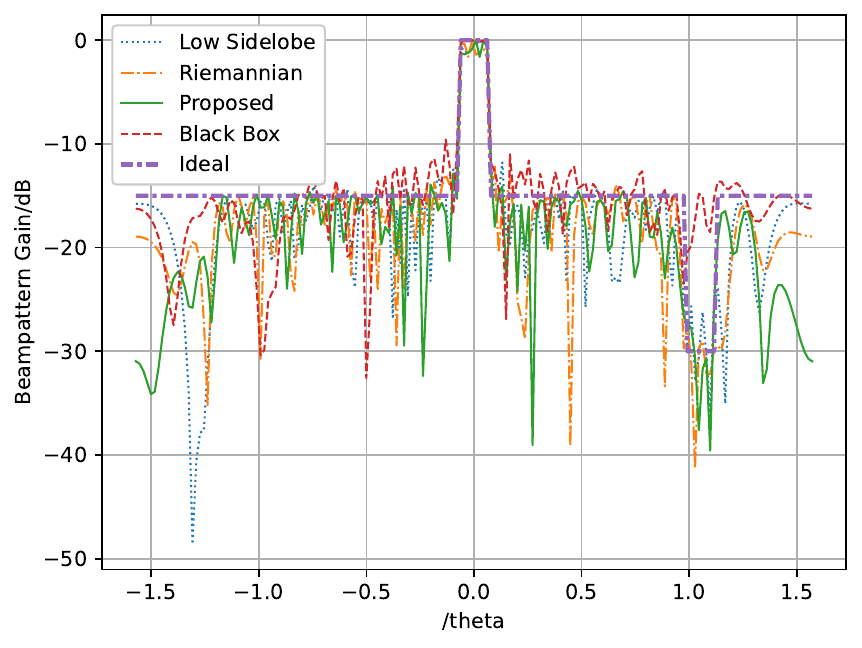}
\caption{The analog beam pattern of single AP (one target and one jammer). }
\label{ana_1T1J}
\end{figure}
Fig. \ref{ana_1T1J} shows the single AP's analog beam pattern gain of different schemes. We choose $[-4^\circ, 4^\circ]$ as the range of mainlobe direction and $[56^\circ, 64^\circ]$ as the range of jamming direction. As is apparent from the results, the performance of the analog beam pattern of the proposed method, assisted by deep unfolding (i.e. benchmarks Low Sidelobe and RGD), is relatively similar. However, their runtime varies, as shown in Table \ref{tab1}. The experimental apparatus employed a 64-bit Intel(R) Core(TM) i7-13700K CPU @3.40GHz with 16GB of RAM. The runtime of the proposed deep unfolding scheme is approximately one third of that of the RGD method and the method in \cite{9611070}. It is evident that implementation of deep unfolding contributes to reduction in computational time required for step size search. The data-driven DNN benchmark, while achieving optimal runtime and mainlobe ripple performance, fails to meet nulling requirements and exhibits higher average sidelobe levels compared to other methods.
\begin{table}[h]
\caption{Single runtime of each scheme}
\begin{center}
\begin{tabular}{|c|c|c|c|c|}
\hline
\textbf{Scheme} & \textbf{{Deep Unfolding}}& \textbf{{RGD}}& \textbf{{Low Sidelobe}} & \textbf{{Black Box}}\\
\hline
\textbf{Runtime}& 1.34s &4.01s &4.56s &0.20s \\
\hline
\end{tabular}
\label{tab1}
\end{center}
\end{table}

\begin{figure}[h]
\centering
\includegraphics[scale=0.5]{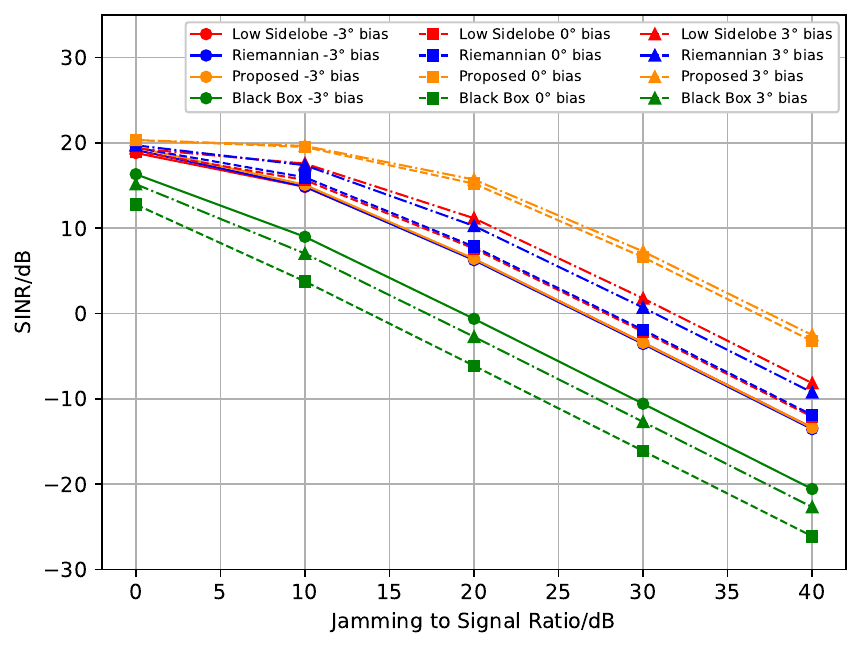}
\caption{Analog beam communication performance under jamming.}
\label{single_AP_SINR}
\end{figure}

Fig. \ref{single_AP_SINR} presents the anti-jamming communication performance of several single-AP analog beam synthesis schemes, where SNR=10 dB. Under the configuration specified in Fig. \ref{ana_1T1J}, we assess the robustness of different schemes against jamming angle biases. The SINR of single UE-k can be expressed as
\begin{equation}
    {\rm SINR}_{k}=\frac{|\mathbf{w}^{\mathrm{BB,H}}\mathbf{W}^{\mathrm{RF,H}}\mathbf{h}_{k}|^2}{\sum_{j=1}^{J}|\mathbf{w}^{\mathrm{BB,H}}\mathbf{W}^{\mathrm{RF,H}}\mathbf{g}_{j}|^2+\sigma^2\Vert\mathbf{w}^{\mathrm{BB,H}}\mathbf{W}^{\mathrm{RF,H}}\Vert^2_2},
\end{equation}
where $\mathbf{h}_{k}$ is the channel of UE-$k$, $\mathbf{g}_{j}$ is the jamming channel of jammer-$j$, and $\sigma^2$ is receiving noise. The results demonstrate that the proposed scheme achieves notably superior anti-jamming performance compared to data-driven approaches, while offering a runtime advantage over conventional model-driven counterparts.

\begin{figure}[h]
\centering
\includegraphics[scale=0.5]{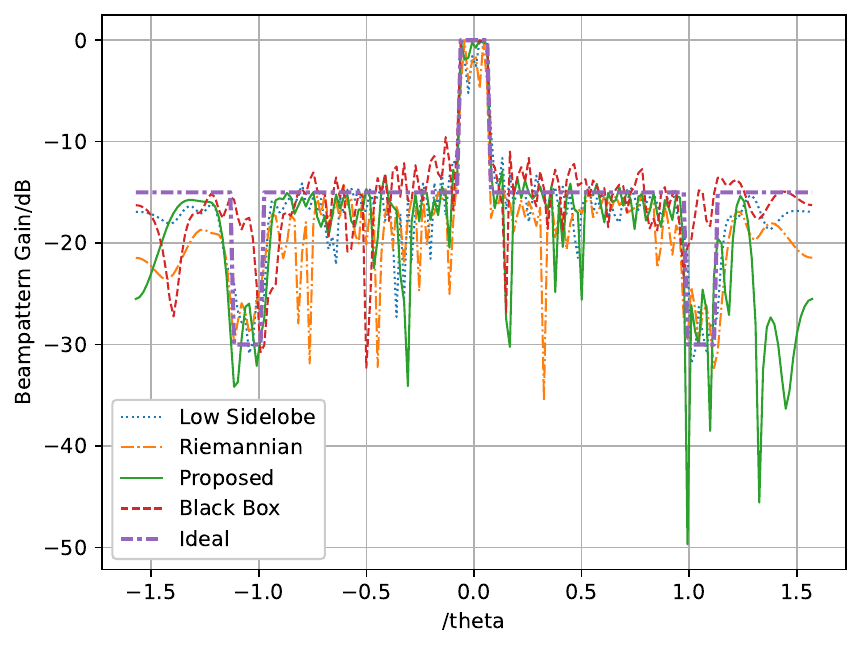}
\caption{The analog beam pattern of single AP (one target and two jammers). }
\label{ana_1T2J}
\end{figure}

Fig. \ref{ana_1T2J} shows simulation with single-AP multi-jamming scenario, where we add the range $[-64^\circ, -56^\circ]$ as another jamming direction. The proposed scheme demonstrates identical mainlobe gain and null performance to the Low Sidelobe baseline in \cite{9611070}. Crucially, while the purely data-driven Black Box approach maintains computational efficiency, it completely fails to generate effective nulls in this dual-jamming scenario, revealing inherent limitations in environmental adaptability and interpretability. The results validate that deep unfolding not only addresses convergence efficiency issues in conventional optimization but also preserves physical interpretability, offering a reliable solution for practical anti-jamming communication systems.

\subsection{CoMP Beam Pattern Performance}

\begin{figure}[h]
\centering
\includegraphics[scale=0.5]{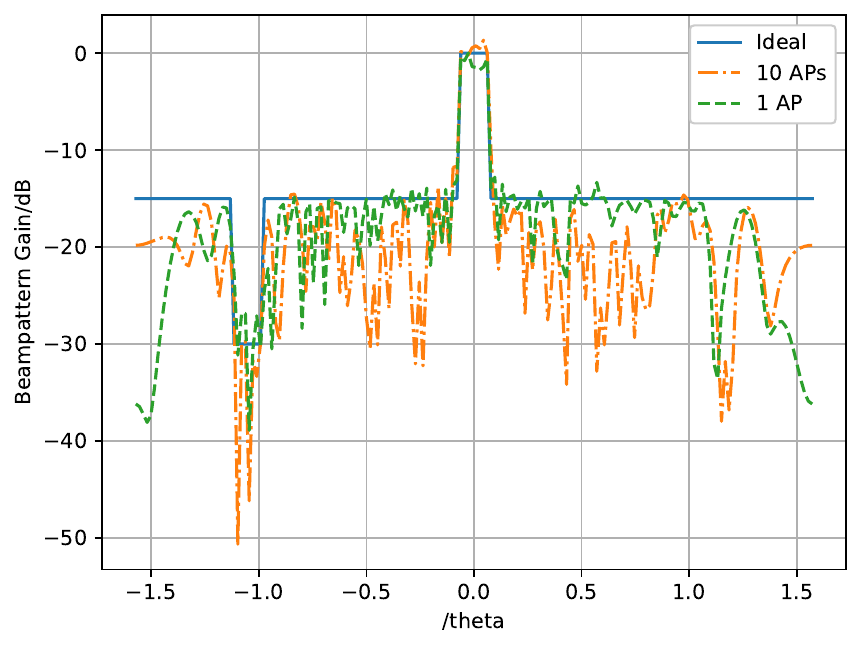}
\caption{The CoMP beam pattern (one target and one jammer). }
\label{COMP1T1J}
\end{figure}

This subsection compares the beamforming performance between CoMP and single-AP systems. Using a designated AP as reference, Fig. \ref{COMP1T1J} shows the beam patterns for mainlobe ($[-4^\circ, 4^\circ]$) and jamming direction ($[-64^\circ, -56^\circ]$). Results demonstrate similar performance between the two schemes under single-target and single-jammer scenarios: comparable mainlobe fluctuations are observed, with the single-AP scheme slightly exceeding the -30dB null depth requirement in some regions, though the overall difference remains insignificant.

\begin{figure}[h]
\centering
\includegraphics[scale=0.5]{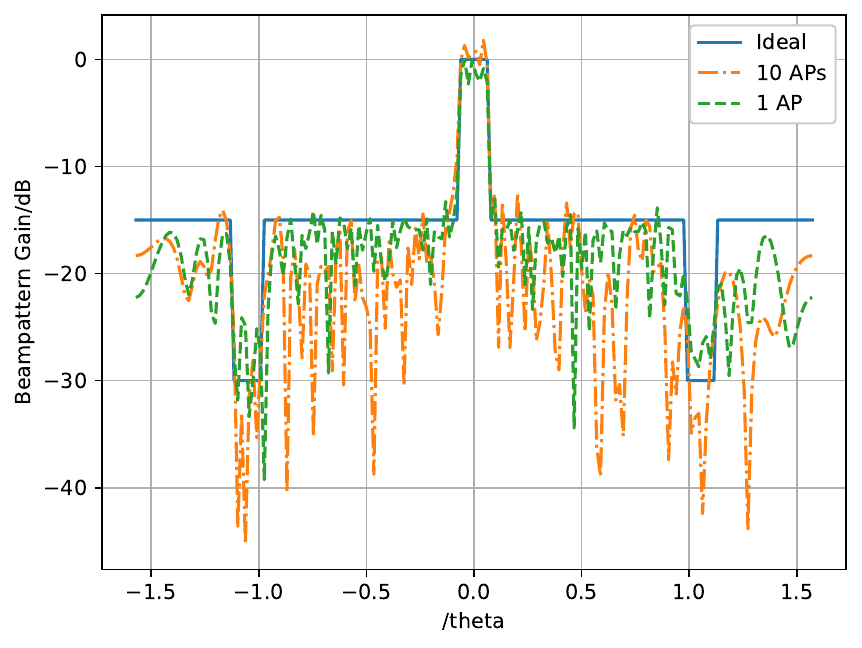}
\caption{The CoMP beam pattern (one target and two jammers). }
\label{COMP1T2J}
\end{figure}

Fig. \ref{COMP1T2J} shows the normalized CoMP beam pattern. When evaluating the performance of CoMP, a reference AP should be selected. We choose $[-4^\circ, 4^\circ]$ as the range of relative mainlobe direction and $[-64^\circ, -56^\circ]\cup[56^\circ, 64^\circ]$ as the range of relative jamming direction. With multiple jamming directions, forming an optimal analog beam is challenging due to the unitary modulus constraint, which limits spatial degrees of freedom. While single-AP analog beamforming shows poorer mainlobe performance than CoMP and cannot simultaneously create wide mainlobes with multiple nulls, CoMP proves decisively superior against multiple jammers, even with just single-RF-chain arrays, demonstrating how coordination overcomes hardware limits.

\begin{figure}[h]
\centering
\includegraphics[scale=0.5]{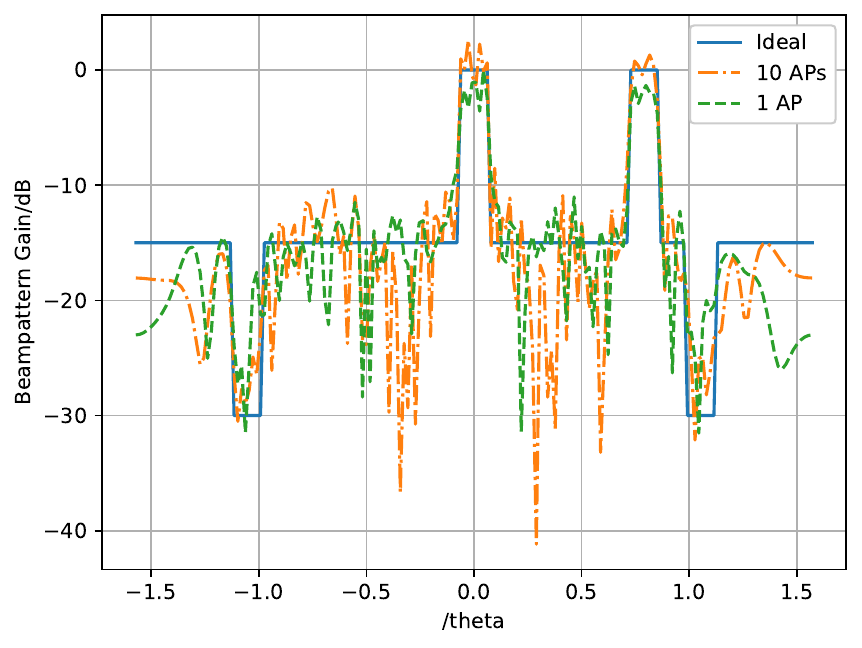}
\caption{The CoMP beam pattern (two target and one jammer). }
\label{COMP2T2J}
\end{figure}

In Fig. \ref{COMP2T2J}, we investigate the performance limits of coordinated beamforming in multi-target multi-jammer scenarios. With two main lobes ($[-4^\circ, 4^\circ]$ and $[41^\circ, 49^\circ]$) and two jammers ($[56^\circ, 64^\circ]$ and $[-64^\circ, -56^\circ]$), results show that while coordinated beamforming outperforms single-AP in mainlobe stability, its performance degrades compared to single-target scenarios due to unit modulus constraints limiting spatial degrees of freedom. Notably, under single RF chain architecture, although improving main lobe performance, some nulls fail to meet the -30dB requirement, revealing inherent limitations in simultaneous multi-mainlobe and deep-null optimization.

\begin{figure}[h]
\centering
\includegraphics[scale=0.5]{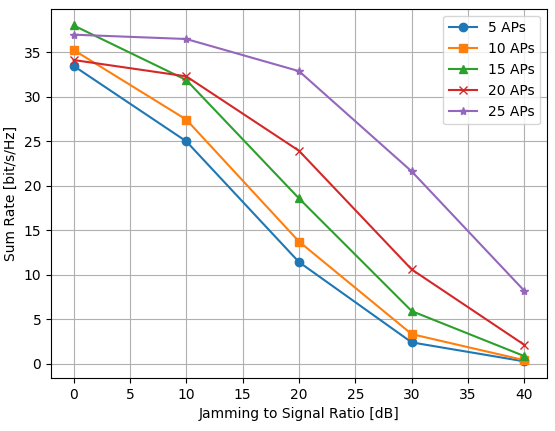}
\caption{Anti-jamming communication performance of DUCoMP-BPS.}
\label{COMP1T1JRmAP}
\end{figure}

We further evaluate the anti-jamming performance of the proposed DUCoMP-BPS scheme through the system sum-rate (5 UEs, SNR=10 dB, with other settings identical to Fig. \ref{COMP1T1J}), as shown in Fig. \ref{COMP1T1JRmAP}. The sum-rate can be expressed as
\begin{equation}
    R = \sum_{k=1}^{K} {\rm log}_{2}(1+{\rm{SINR}_{k}}).
\end{equation}
As the number of APs increases, the sum-rate under the same jamming level gradually improves, which is particularly pronounced at high jammer-to-signal ratios. This is because a larger number of APs enables the formation of beams with higher precision and stronger gains.

\subsection{Computational Complexity and Scalability Analysis}

\begin{figure}[h]
\centering
\includegraphics[scale=0.6]{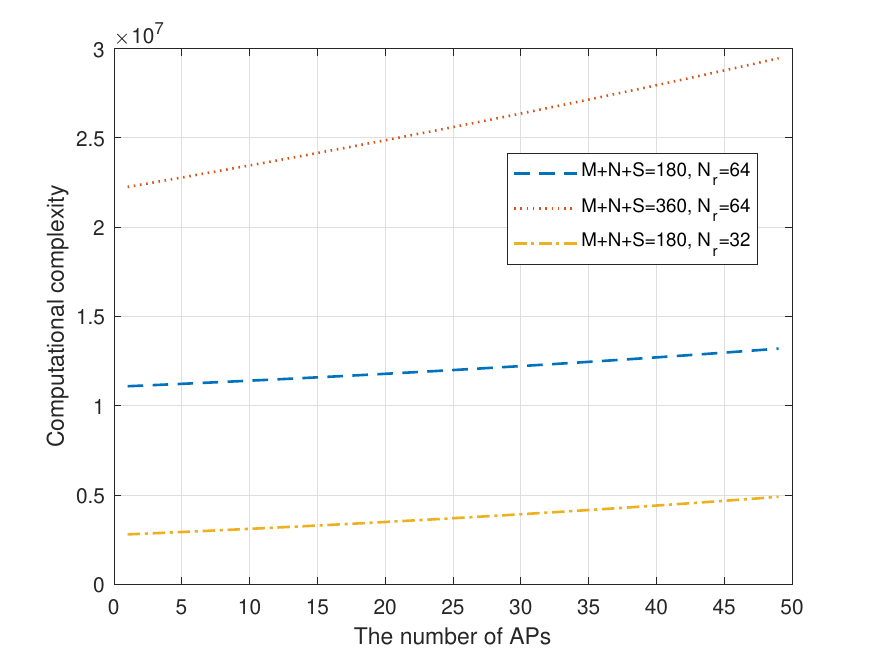}
\caption{The complexity of proposed scheme. }
\label{Complexity}
\end{figure}

We finally evaluate the computational complexity through simulations. Fig. \ref{Complexity} shows near-linear complexity scaling with the number of APs as mentioned in Section \ref{complex_ref}, which demonstrates good scalability for CF MIMO systems.

\section{Hardware Acceleration and Architecture Design}

Building upon the simulation-based validation in the previous section, this section focuses on translating the proposed AI accelerated analog beam algorithm in DUCoMP-BPS scheme into a deployable hardware solution. We detail the design of a heterogeneous ARM plus FPGA architecture that accelerates AI-assisted beamforming, the allocation of computational tasks between hardware and software, and the optimization of neural network inference for real-time operation. These hardware-level enhancements serve as a bridge between algorithmic simulation and practical field validation, thereby ensuring that the proposed scheme is both computationally efficient and ready for real-world deployment.

\subsection{Top-level Design of Hardware Platform for AI Acceleration}

\begin{figure}[h]
\centering
\includegraphics[scale=0.4]{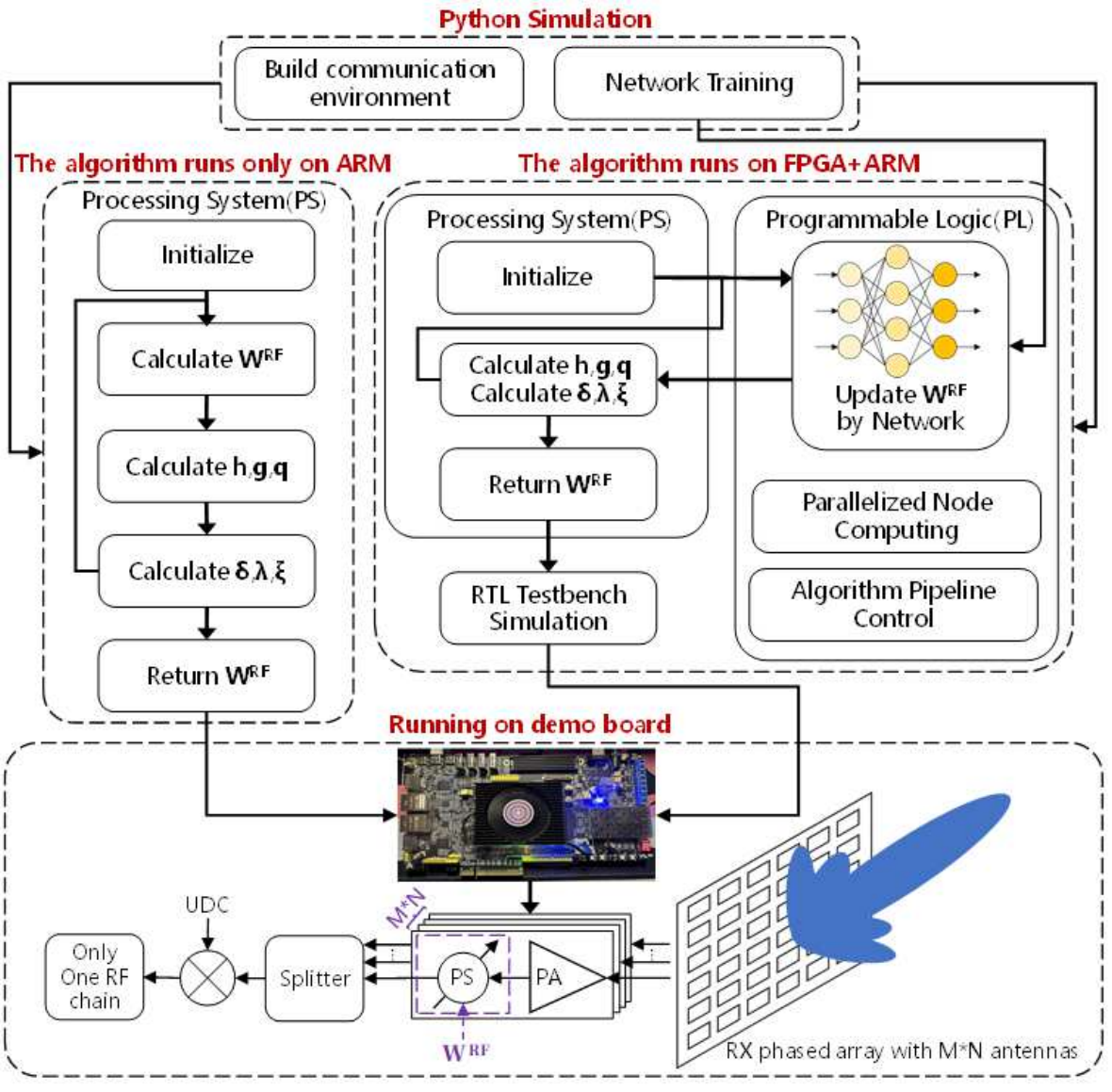}
\caption{System-level design and hardware build.}
\label{sys_des}
\end{figure}

The top-level architecture of the AI acceleration platform designed in this paper is shown in Fig. \ref{sys_des}\cite{liu2024analog}. We compared the original algorithm without AI acceleration with the AI-accelerated algorithm designed in this paper on the demo board, evaluating their full operational workflows under two heterogeneous schemes. This design employs a hierarchical approach integrating software and hardware, encompassing the entire process from software-level communication environment construction, simulation, neural network design, and network training, to hardware-level architecture design, network mapping, and practical deployment verification.

At the software level, the communication system environment and neural network design were constructed using Python to complete system simulation and neural network training. At the hardware level, two distinct paths emerged based on algorithm performance comparisons: the original algorithm runs entirely on the ARM processor within the Processing System (PS), performing parameter calculations required for beamforming; deploying the AI-accelerated algorithm on a heterogeneous computing architecture integrating ARM and FPGA. Here, the ARM handles control and scheduling, while computationally intensive tasks—specifically hyperparameter prediction—are deployed on the FPGA within the Programmable Logic (PL). Acceleration is achieved through node parallel computing design and pipeline design. Both approaches rely on the AXI bus for data exchange between PS and PL, with functional verification conducted via register-transfer level (RTL) simulation and hardware testing.

This study further employs the High-Level Synthesis (HLS) approach to achieve dynamic reconfiguration of hardware modules, enabling adaptation to future algorithmic variants and extensions. The final deployment on the Xilinx Zynq-7100 series platform validated the significant computational efficiency advantages of the AI acceleration architecture investigated herein.

\subsection{Hardware Mapping and Acceleration Design for Neural Network Computing}

Considering the balance between the resources and performance of FPGA, the algorithm five-layer neural network is simplified to three-layer neural network, the output activation function of the first and second layer is CReLU function, the output of the last layer is the sum of the real part and the imaginary part, and in order to ensure the non-negativity of the output step parameter, the activation function of the last layer is set to the absolute value (Abs) function. 

\begin{figure}[h]
\centering
\includegraphics[scale=0.35]{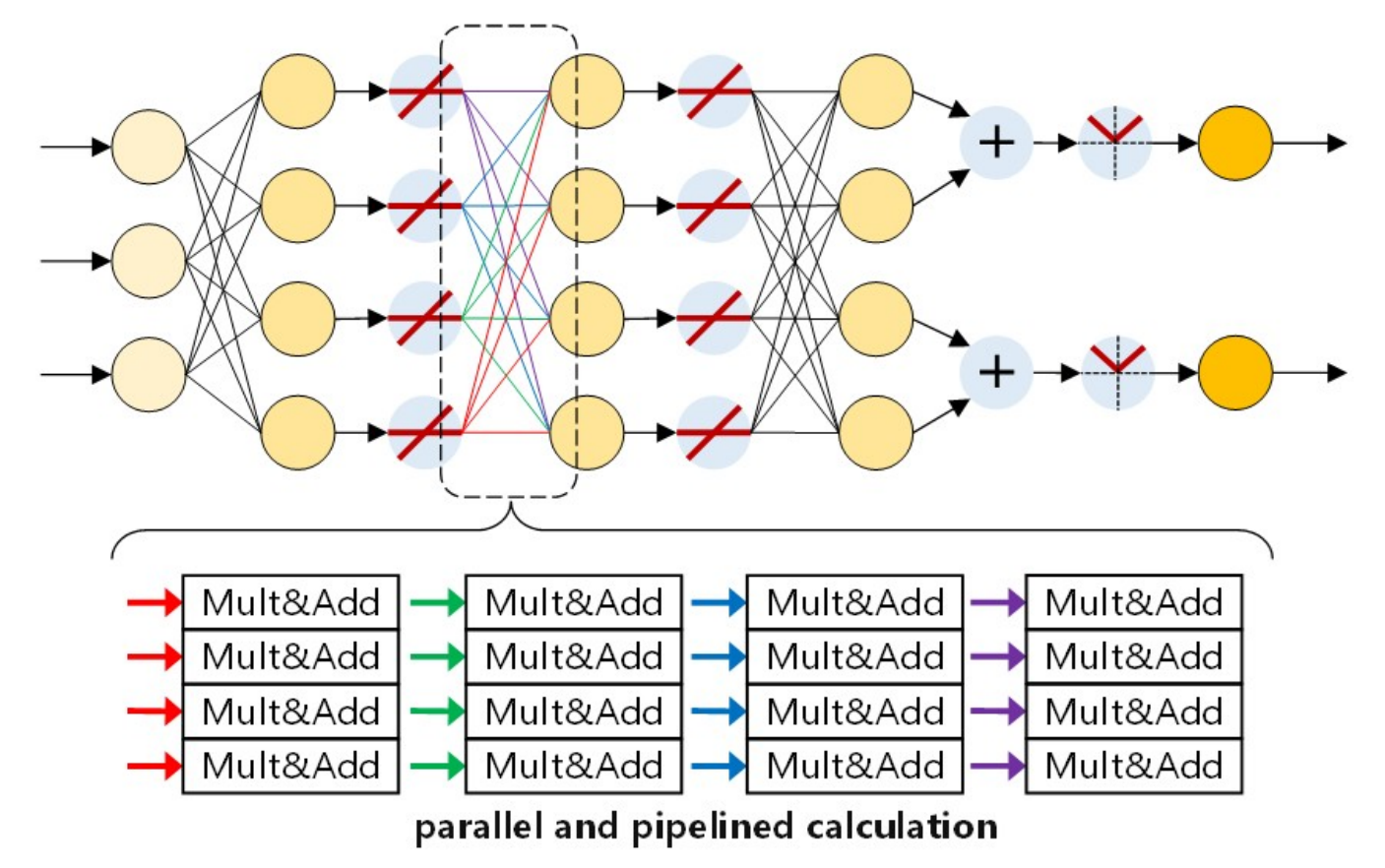}
\caption{Hardware mapping and acceleration design for neural network computing.}
\label{hardmod}
\end{figure}

The hyperparameter complex neural network design is characterized by the following:
\begin{itemize}
    \item Complex linear layer constitutes the backbone of feature extraction in the complex domain, decouples complex multiplication into real multiplication by real-imaginary partial solution strategy, and at the same time realizes parallel execution paths, which is adapted to the parallel pipeline architecture of FPGA programmable logic.
    \item CReLU is used to replace the traditional ReLU, and its output is cascaded through Sum and Absolute Value layers, which reduces the resource consumption while maintaining the feature expressiveness.
    \item The FPGA dynamically adjusts the balance between parallelism, pipelining levels, and resource utilization as needed. Fig. \ref{hardmod} illustrates the hardware mapping and acceleration design for neural network computing, showcasing hardware parallel computation and pipelining mechanisms between layers. Compared to the limited parallel capabilities of ARM, this design achieves high-performance neural network computation acceleration.

\end{itemize}


Our FPGA design features a highly parallel architecture with 128 floating-point lanes and a deeply pipelined core with 11-cycle initial latency, achieving 128 32-bit floating-point operations per cycle once the pipeline is full. Table \ref{tab2} shows the specific resource consumption of AI acceleration algorithms.
\begin{table}[h]
\caption{Hardware Resource Consumption of AI Acceleration Algorithms}
\begin{center}
\begin{tabular}{|c|c|c|c|c|}
\hline
\textbf{Resource} & \textbf{{LUT}}& \textbf{{FF}}& \textbf{{BRAM}} & \textbf{{DSP}}\\
\hline
\textbf{Consumed}& 144676 &206131 &40 &482 \\
\hline
\textbf{Total}& 277400 &554800 &755 &2020 \\
\hline
\end{tabular}
\label{tab2}
\end{center}
\end{table}

\subsection{Hardware Simulation and Analysis}

\begin{figure}[h]
\centering
\includegraphics[scale=0.45]{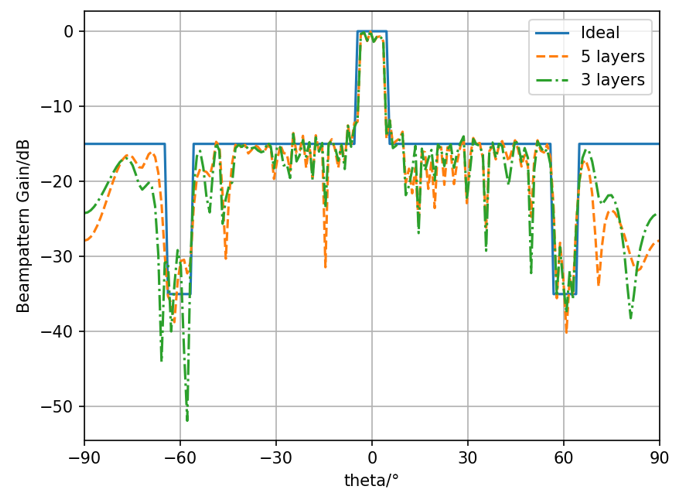}
\caption{Analog beam patterns of 5 and 3 complex linear layers.}
\label{35ab}
\end{figure}

In order to realize the AI-assisted acceleration of the proposed scheme in an ARM+FPGA heterogeneous hardware platform, this study simplifies the original 5-layer hyperparameter prediction neural network into a 3-layer structure. The 3-layer model achieves a 65\% reduction in logic resources and a 97\% reduction in on-chip memory utilization on the FPGA side. The 3-layer network model exhibits a high degree of concordance with the 3-layer network model with respect to the key indexes of the mainlobe position, the depth of the null, and the level of the sidelobes, as illustrated in Fig. \ref{35ab}. However, a divergence emerges in terms of the average convergence time. It is evident that the two phenomena coincide with each other. However, it is only in the average convergence time that a 100\% training delay is observed in comparison to the 5-layer network model.


\begin{figure}[h]
\centering
\includegraphics[scale=0.45]{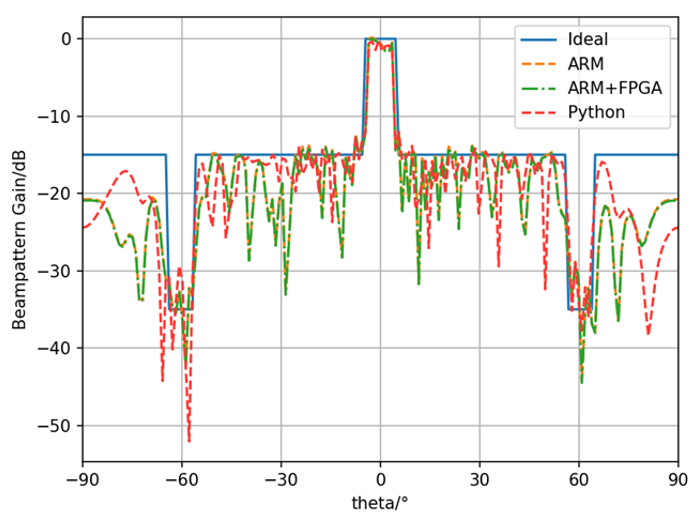}
\caption{Analog beam patterns of ARM and ARM+FPGA.}
\label{armfpga}
\end{figure}

The trained 3-layer inference network is integrated into the heterogeneous hardware platform framework, and the regularized hyperparameters are generated in real time by network parallel accelerated computation and pipeline optimization. The runtime of ARM+FPGA is 92${\mu}$s while the runtime of ARM is 1685${\mu}$s, which shows that the combination of ARM and FPGA reduces the computation time consumed for beam weights by 94.5\% compared to the non-accelerated scheme on ARM. Meanwhile, the hardware simulation direction graphs (yellow and green lines) and the software simulation results (red line) demonstrate a high degree of correlation with regard to the key indexes such as the mainlobe position, null depth, and sidelobes level, as illustrated in Fig. \ref{armfpga}.

\section{Practical Measurement and Real-World Validation}

Following the hardware acceleration and architecture design in the previous section, this section moves from FPGA-based deployment to real-world verification. We establish a calibrated measurement process in a controlled anechoic chamber to evaluate the AI-accelerated analog beamforming algorithm under practical hardware conditions. By modeling and compensating for key non-idealities, such as channel gain/phase inconsistency, phase quantization, and nonlinear phase mapping, we ensure that the beam pattern synthesis results reflect the actual performance of the proposed DUCoMP-BPS scheme. This hardware-in-the-loop measurement serves as the final validation step before potential field deployment.

\subsection{The Antenna Channel Error Modeling of Uniform Planar Array (UPA)}
For a UPA, we take an $M \times N$ array as an example and assume that the ideal weight of the $(m,n)$ antenna channel is $W_{mn}$. Thus, the ideal beam pattern can be expressed as
\begin{equation}\label{UPA}
    F\left( {\theta ,\varphi } \right) = \sum\limits_{m = 1}^M {\sum\limits_{n = 1}^N {{W_{mn}}{e^{j\left( {mu + nv} \right)}}} } ,
\end{equation}
where $u=\sin \theta \sin \varphi $ and $v=\cos \theta$ are the virtual angles of UPA.

Array antennas exhibit non-uniformity in gain and phase across each antenna channel due to issues such as printed circuit board (PCB) manufacturing processes, chip manufacturing processes, and impedance errors. Assuming that the amplitude and phase of each antenna channel on the same array are uncorrelated, and that the amplitude and phase of each antenna channel follow the same distribution, as
\begin{equation}
\begin{array}{l}
     \delta _{mn}^A \sim N\left( {0,\sigma _a^2} \right),  \delta _{mn}^P \sim N\left( {0,\sigma _p^2} \right), \\
     {m = 1,2, \cdots ,M;n = 1,2, \cdots ,N} ,
\end{array}
\end{equation}
where $\delta _{mn}^A$ and $\delta _{mn}^P$ represent the amplitude and phase non-consistency error of the $(m,n)$ antenna channel, respectively. Then, the actual weight of the $(m,n)$ antenna channel is 
\begin{equation}
    W_{mn}^D = \left( {1 + \delta _{mn}^A} \right){e^{j\delta _{mn}^P}}{W_{mn}}.
\end{equation}

For a practical analog phased array, the phased shifter normally has limited accuracy. The minimal quantization accuracy is ${\Delta _b} = 2\pi /{2^b}$ for a $b$ bit phased shifter\cite{9648341}, \cite{10333584}. Considering the unitary modulus constraint of analog phased arrays and assuming the ideal phase configuration is $\phi _{mn}$, the phase after quantization of the weight ${W_{mn}}$ can be expressed as
\begin{equation}
    {W_{mn}} = {e^{j\left[ {\frac{{{\phi _{mn}}}}{{{\Delta _b}}}} \right]{\Delta _b}}},
\end{equation}
of which the phase quantization error is 
\begin{equation}
    {\phi _{quant\_err}} = {\phi _{mn}} - \left[ {\frac{{{\phi _{mn}}}}{{{\Delta _b}}}} \right]{\Delta _b},
\end{equation}
and the actual quantized weight of the $(m,n)$ antenna channel is 
\begin{equation}
    W_{mn}^D = \left( {1 + \delta _{mn}^A} \right){e^{j(\left[ {\frac{{{\phi _{mn}}}}{{{\Delta _b}}}} \right]{\Delta _b} + \delta _{mn}^P)}}.
\end{equation}

In practical measurement, we could specifically design the weight to compensate these errors. From the perspective of antenna weight design, its non-consistency error can be obtained through measurements in an anechoic chamber and corrected. Considering the computational complexity, the error of phase quantization is commonly measured using cosine similarity to evaluate the consistency of phase vectors before and after quantization in the complex vector space. The impact of phase quantization error can be reduced by finding the quantization weight with the highest cosine similarity.

\subsection{Measurement And Analysis}
\begin{figure}[h!]
\centering
\includegraphics[scale=0.55]{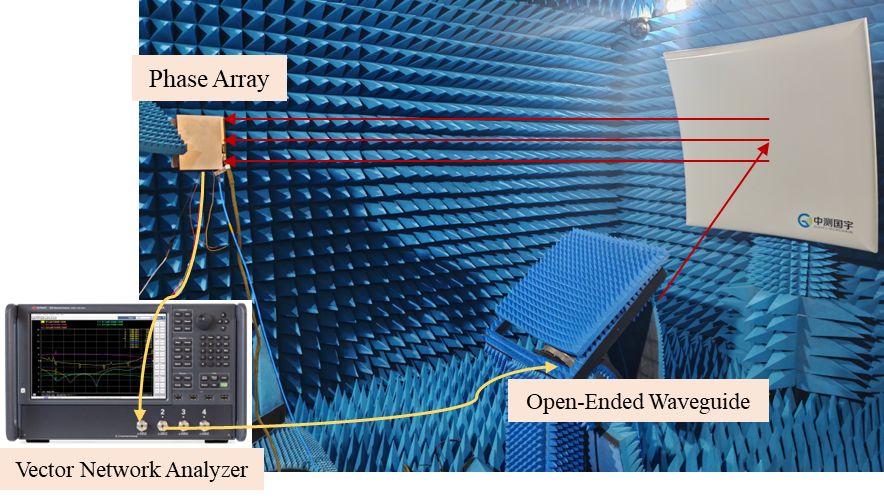}
\caption{The illustration of the measurement environment.}
\label{meas_1}
\end{figure}

To test the actual beam pattern performance generated by the algorithm, we set up an anechoic chamber test environment as shown in Fig. \ref{meas_1}. The single-tone signal source is placed at the focal point of the parabolic reflector, where the spherical waves it emits are converted into plane waves by the reflector. These parallel electromagnetic waves are incident on the phased array antenna under test, ensuring that the signals received by the array meet far-field conditions, thereby enabling precise beam pattern measurements. The phased array under test consists of 8x16 elements, with phase shifter control accuracy of 6 bits. The single-tone signal source operates at a frequency of 26.75 GHz with a transmission power of 30 dBm. The rotation platform supporting the phased array has an angular accuracy of less than $0.05^\circ$.

\begin{figure}[htbp]
\centering
\subfloat[]{\includegraphics[scale=0.18]{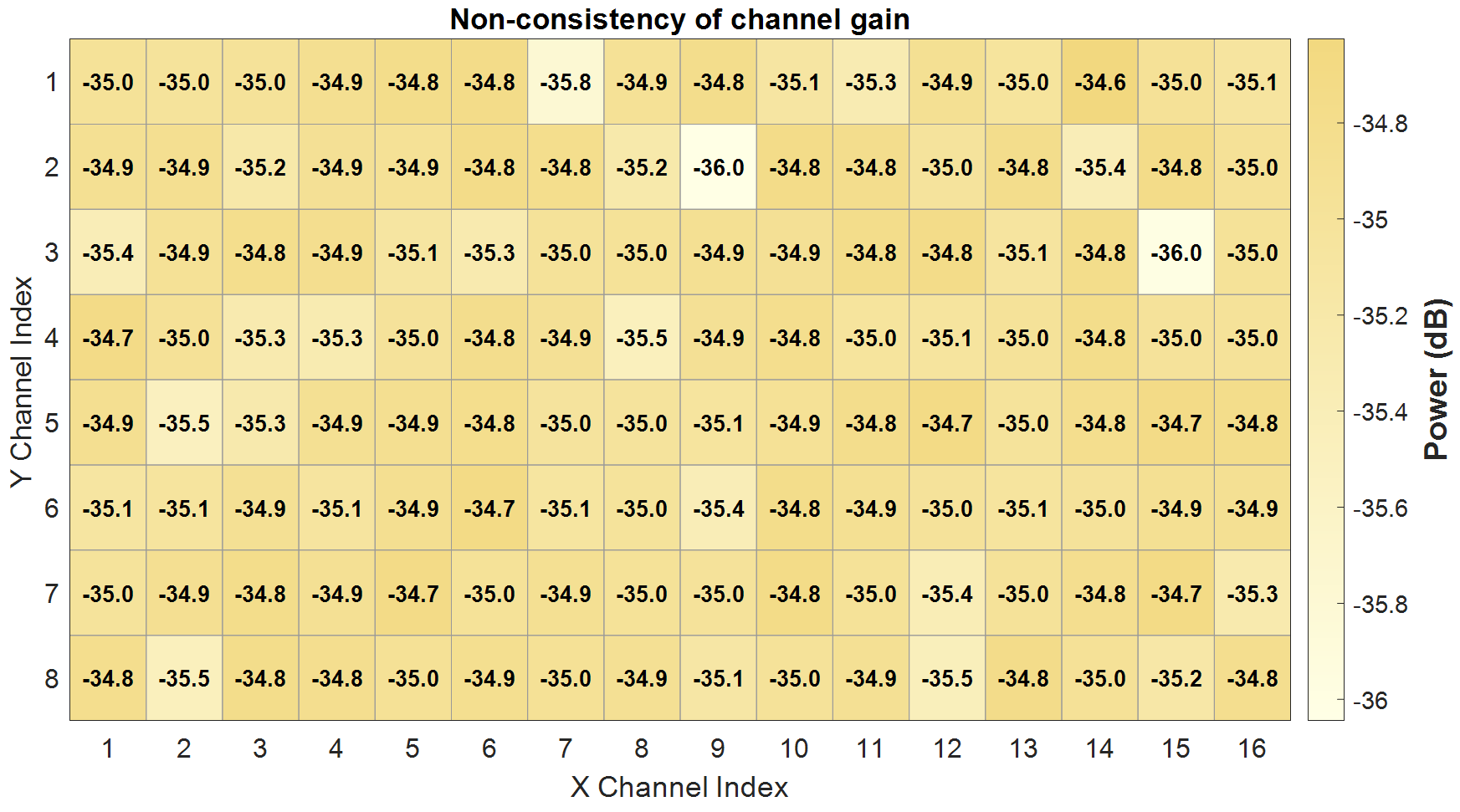}\label{meas_2}}\\
\subfloat[]{\includegraphics[scale=0.18]{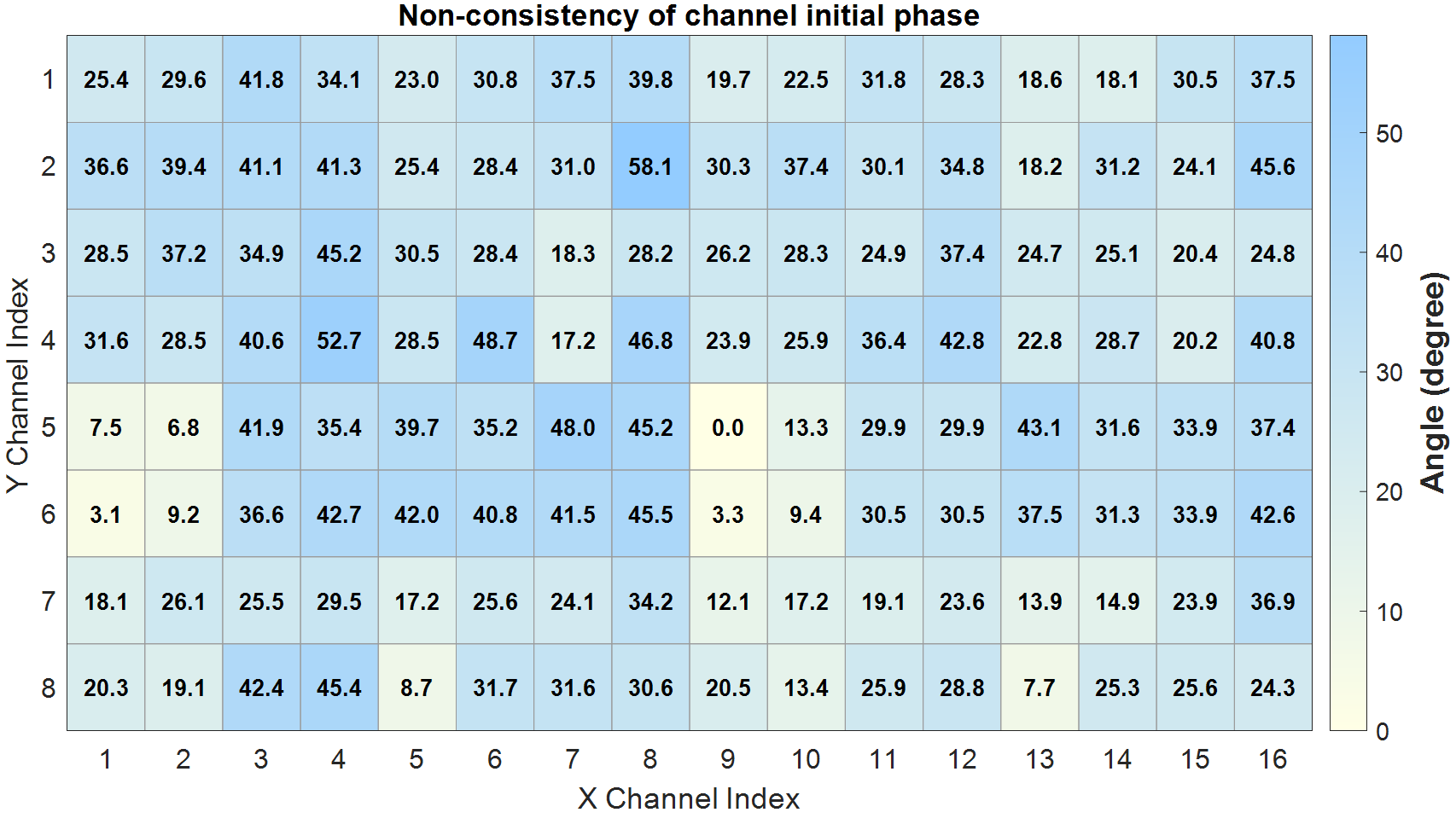}\label{meas_3}}
\caption{(a) The non-consistency error of antenna channel gain. (b) The non-consistency error of antenna channel initial phase.}
\end{figure}

We test each channel individually, collect spectrum analyzer data, and obtain the channel gain inconsistency and channel initial phase inconsistency results, as shown in Fig. \ref{meas_2} and Fig. \ref{meas_3}. During the algorithm design process, the measured inconsistency error will be used as a substitute.



\begin{figure}[h!]
\centering
\includegraphics[scale=0.30]{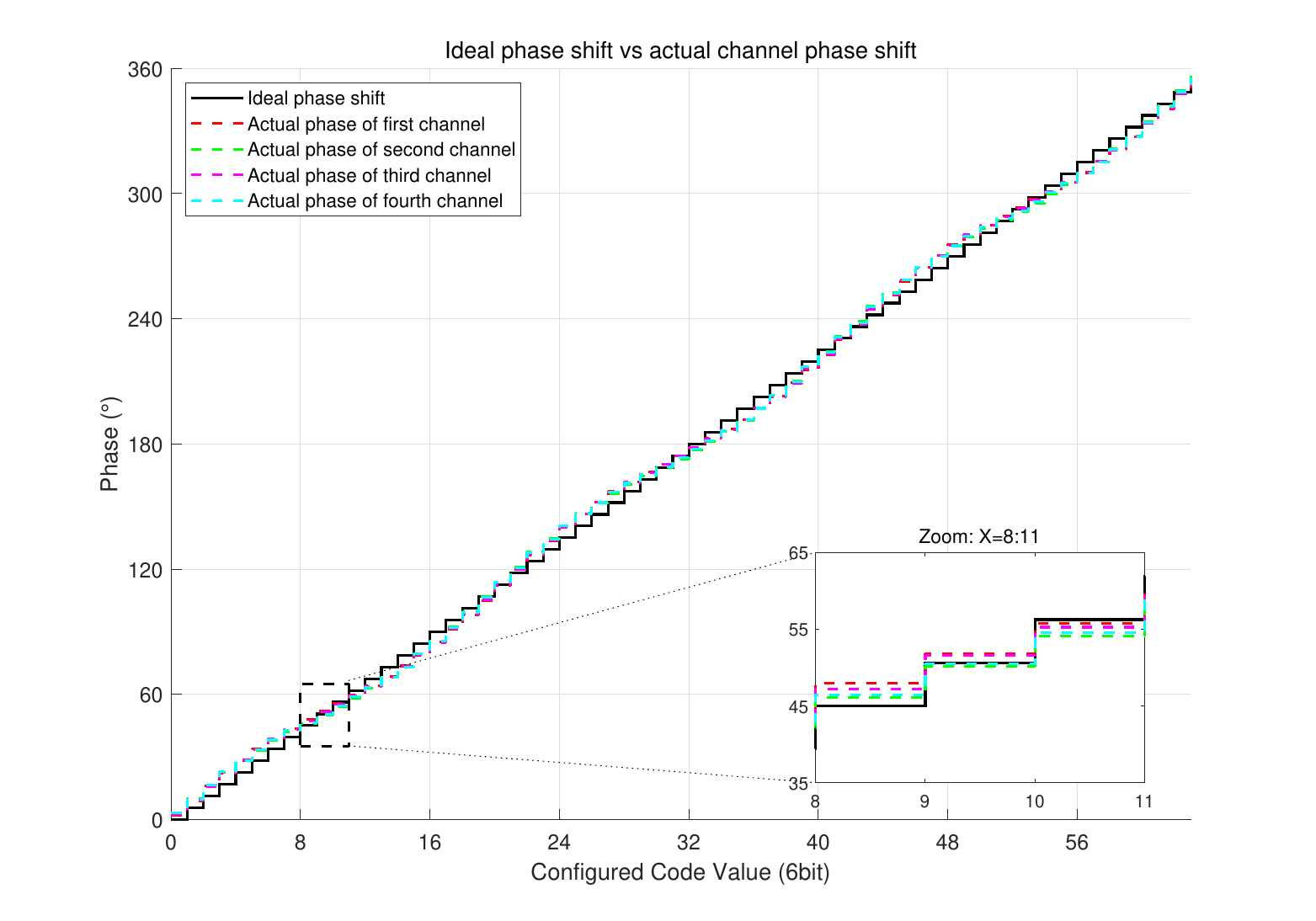}
\caption{The illustration of non-linear mapping of multi-channel phase shift.}
\label{meas_4}
\end{figure}

After eliminating the initial phase inconsistency of the channels, the quantization modeling and errors of the phase are generally considered, such as a 6-bit phase shifter, where the phase shifter configuration code values from 0 to 64 are typically linearly mapped to 0° to 360°. In reality, phase inconsistency in phased arrays also manifests as a nonlinear mapping relationship between the phase configuration values and actual values of each channel, meaning that phase errors must also account for nonlinear effects, with error magnitudes comparable to those of quantization errors. By sequentially enabling each channel of the phased array while iterating through each phase shifter configuration code value, we obtain the nonlinear mapping results of multi-channel phases as shown in Fig. \ref{meas_4}. The results are substituted for linear mapping using a lookup table, while considering the effects of quantization, and incorporated into the calculation of cosine similarity.

\begin{figure}[h!]
\centering
\includegraphics[scale=0.30]{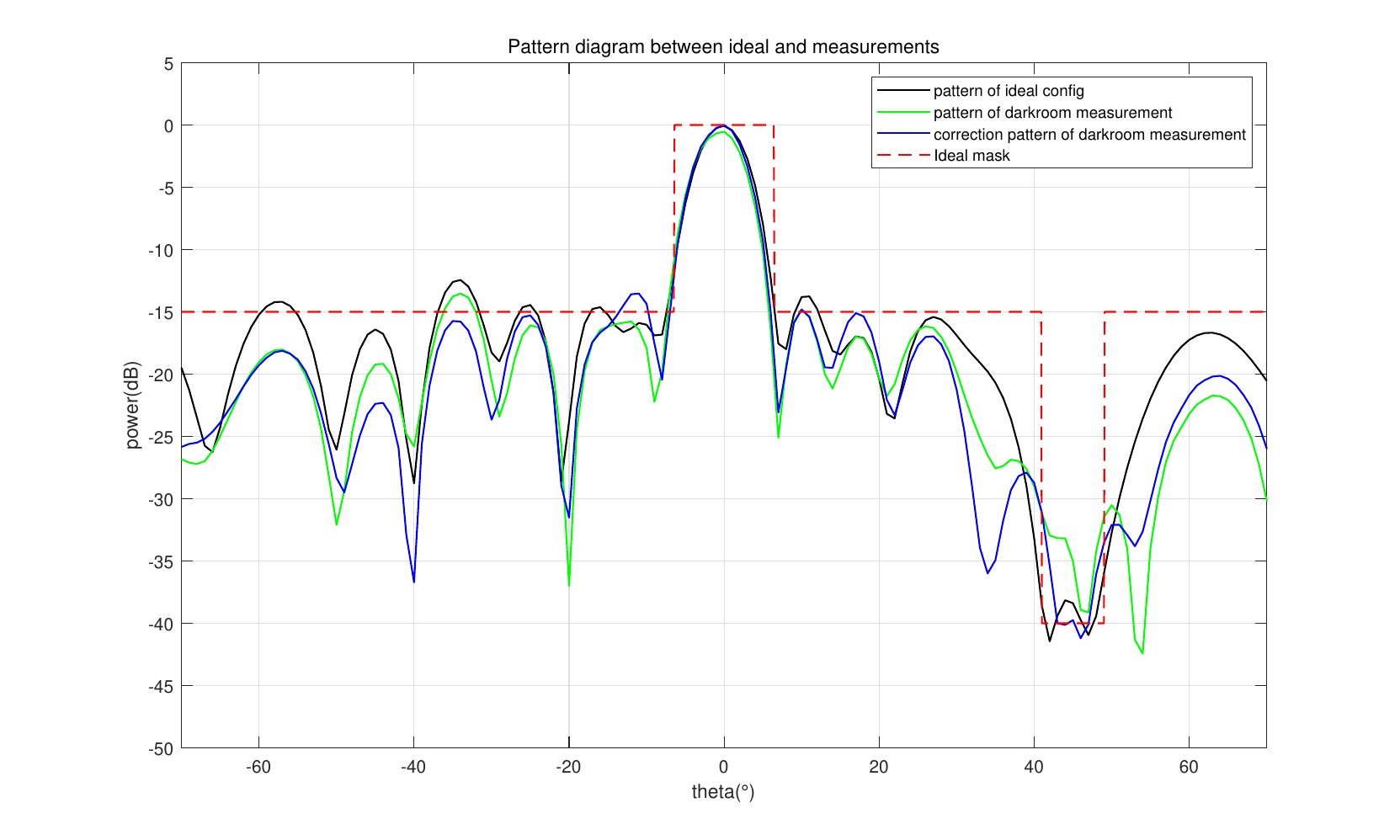}
\caption{The illustration of measured beam pattern and ideal one.}
\label{meas_5}
\end{figure}

Set the single target direction to horizontal $0^\circ$ and elevation $0^\circ$, the single jamming direction to horizontal $45^\circ$ and elevation $0^\circ$, the side lobe suppression level to -15 dBm, the null suppression level to -40 dBm, and the null width to $8^\circ$. Considering two scenarios—the linear mapping relationship of phase configuration and the nonlinear mapping relationship corrected based on channel measurements—we conducted anechoic chamber tests on the beamforming weights designed in this algorithm, yielding the directional pattern results shown in Fig. \ref{meas_5}. As shown in the figure, the green line represents the uncorrected directional pattern, which is more severely distorted due to phase errors caused by nonlinear mapping. The measured directional pattern still deviates slightly from the theoretical value because, while this paper corrects the main non-ideal factors of the phased array, other non-ideal factors have not been considered, and perfect zero-error correction is not achievable. The measured directional pattern has basically achieved the design objectives of the algorithm, verifying its effectiveness.

\section{Conclusion}
This paper proposed a DUCoMP-BPS scheme to address the high complexity, poor adaptability, and scalability limitations of traditional cell-free anti-jamming beamforming. In this design, APs independently perform analog beamforming based on local angle information, while the CPU optimizes digital beamforming with only a single AP–CPU interaction, thereby reducing fronthaul overhead. A deep unfolding-based step size optimization with CvNN further accelerates analog beamforming, achieving up to 67\% runtime reduction compared with conventional methods while delivering superior nulling and sidelobe suppression over data-driven baselines. The scheme scales linearly with the number of APs, and its hardware feasibility is validated on an ARM–FPGA heterogeneous platform with efficient, low-latency execution. Finally, anechoic chamber measurements under hardware imperfections confirm robust performance, demonstrating the strong potential of DUCoMP-BPS for practical deployment.

\bibliographystyle{ieeetr}  
\bibliography{ref.bib}  



\end{document}